\documentclass[aps,prc,superscriptaddress,showpacs,amssymb,amsmath,amsfonts,twocolumn,floatfix]{revtex4-1}
\usepackage{graphicx}
\newcommand*{\GWU}{Data Analysis Center at the Institute for Nuclear Studies,\\
        Department of Physics\\
        The George Washington University, Washington, D.C. 20052}

\newcommand{\en}{\ensuremath{\eta N}}
\newcommand{\gn}{\ensuremath{\gamma N}}
\newcommand{\gnpn}{\ensuremath{\gamma N\to \pi N}}
\newcommand{\Kbar}{\ensuremath{\overline{K}}}
\newcommand{\pD}{\ensuremath{\pi \Delta}}
\newcommand{\pn}{\ensuremath{\pi N}}
\newcommand{\pnten}{\ensuremath{\pi N\!\to\!\eta N}}
\newcommand{\pntpn}{\ensuremath{\pi N\!\to\!\pi N}}
\newcommand{\ppn}{\ensuremath{\pi\pi N}}
\newcommand{\rn}{\ensuremath{\rho N}}

\begin{document}

\title{Unified Chew-Mandelstam SAID analysis of pion photoproduction data}
\author{Ron L.\ Workman} 
\affiliation{\GWU}
\author{Mark W.\ Paris}
\altaffiliation[Present address: ]{Theoretical Division, 
  Los Alamos National Laboratory, Los Alamos, NM 87545, USA}
\affiliation{\GWU}
\author{William J.\ Briscoe}
\affiliation{\GWU}
\author{Igor I.\ Strakovsky}
\affiliation{\GWU}

\date{\today}

%%%%%%%%%%%%%%%%%%%%%%%%%%%%%%%%%%%%%%%%%%%%%%%%%%%%%%%%%%%%%%%%%%%%% 
\begin{abstract}
A unified description of single-pion photoproduction data, together with
pion- and eta-hadroproduction data, has been achieved in a Chew-Mandelstam 
parametrization which is consistent with unitarity at the two-body level. 
Energy-dependent and single-energy partial wave analyses of pion 
photoproduction data have been performed and compared to previous SAID 
fits and multipoles from the Mainz and Bonn-Gatchina groups.
\end{abstract}
%%%%%%%%%%%%%%%%%%%%%%%%%%%%%%%%%%%%%%%%%%%%%%%%%%%%%%%%%%%%%%%%%%%%%%


\maketitle

%%%%%%%%%%%%%%%%%%%%%%%%%%%%%%%%%%%%%%%%%%%%%%%%%%%%%%%%%%%%%%%%%%%%%%%%%%
\section{Introduction}
\label{sec:intro}

A wealth of $\gnpn$ data, for single- and double-polarization observables, 
is anticipated from electromagnetic facilities worldwide over the coming
months and years. These data will be pivotal in determining the underlying
amplitudes in complete experiments, and in discerning between various 
microscopic models of multichannel reaction theory. 

The focus of precision electromagnetic measurements, over the nucleon resonance 
region, is to more fully map the non-perturbative regime of quantum
chromodynamics, the fundamental theory of the strong interaction, to
shed light on its confining and chiral symmetry breaking properties.
These electromagnetic data take the field to 
the next and necessary level of precision. This is required in order
to obtain a theoretical description of the nucleon that both explains and 
subsumes the simple constituent quark model, which has provided a
qualitative picture of nucleon structure and reactions. 
The expected data heralds an era of precision
hadron spectroscopy, particularly for baryons, and has ushered in a
renaissance in hadronic reaction theory. Significant refinements in
the quality and quantity of available data offer the opportunity to
develop more sophisticated models of hadronic reactions, constrained
by fundamental principles of field theory, such as unitarity and gauge
invariance, which have model dependencies under better control, if not
eliminated. Such a complete and successful phenomenology would appear 
to be a prerequisite for a deeper understanding in terms of quarks 
and gluons~\cite{Pennington:2011bq}.

The present manuscript details multipole analyses of the single-pion
photoproduction data using a parametrization form related to, but
an improvement upon, previous SAID
parametrizations~\cite{Arndt:1989ww,said2002,Workman:2011vb}. The 
energy-dependent (ED) analysis is performed over the center-of-mass 
energy ($W$) range from the near-threshold region to about 2.5~GeV, 
including resonances through the fourth resonance region. We also 
generate single energy solutions (SES), which fit the data over narrow 
energy bins assuming phase information obtained from the ED solutions. 
The relations between ED and SES fits have been extensively studied in 
Ref.~\cite{Workman:2011hi}. A detailed discussion of amplitude and 
observable conventions is also given in this source. 

The fitted pion-photoproduction database is identical to that used in 
our most recent~\cite{Workman:2011vb} SN11 analysis, based on the 
standard SAID parametrization. In the following section, we compare 
the previous and present SAID fit forms used to analyze these data. 
Extracted multipoles are compared to previous SAID fits, and those 
from other groups, in Sec.~\ref{sec:mpole}. Our results and their 
implications are summarized in Sec.~\ref{sec:sum}.

%%%%%%%%%%%%%%%%%%%%%%%%%%%%%%%%%%%%%%%%%%%%%%%%%%%%%%%%%%%%%%%%%%%%%%%
\section{Formalism}
\label{sec:form}

The Chew-Mandelstam (CM) energy-dependent (ED) parametrization for the
hadronic $T$ matrix, described in Ref.~\cite{Arndt:2006bf}, has been
used in a recent coupled-channel fit of $\pn$ elastic scattering
and $\pn\to\en$ reaction data.  It gives a realistic description of the
data with $\chi^2$ per datum better than any other parametrization or
model, to our knowledge~\cite{beij09,Paris:2010tz}.  The parametrization 
form used in this fit is given as
\begin{align}
\label{eqn:Tab}
	T_{\alpha\beta} &= \sum_\sigma [1-\Kbar C]^{-1}_{\alpha\sigma}
	\Kbar_{\sigma\beta} ,
\end{align}
where $\alpha$, $\beta$, and $\sigma$ are indices for the 
considered channels, $\pn,\pD,\rn$, and $\en$. This parametrization 
has been discussed in Refs.~\cite{Arndt:1985vj,Arndt:1995bj,Arndt:2006bf}.
Given the success of this approach in the hadronic two-body sector, its
application to the study of meson photoproduction is warranted. The main 
result of the present study is the use of the information encoded in 
Eq.~\eqref{eqn:Tab} by employing the factor $[1-\Kbar(W)C(W)]^{-1}$ 
(called the ``hadronic rescattering matrix'') in the photoproduction 
parametrization form.

The CM form of Eq.~\eqref{eqn:Tab} may be extended to include the 
electromagnetic channel as:
\begin{align}
\label{eqn:Tag}
	T_{\alpha\gamma} &= \sum_\sigma [1-\Kbar C]^{-1}_{\alpha\sigma}
	\Kbar_{\sigma\gamma} .
\end{align}
Here, $\gamma$ denotes the electromagnetic channel, $\gn$, and
$\sigma$ denotes the hadronic channels which appear in the
parametrization of the hadronic rescattering matrix, $[1-\Kbar
C]^{-1}$. Note that by sharing the common factor, $[1-\Kbar
C]^{-1}$ which encodes, at least qualitatively speaking, the 
hadronic channel coupling (or rescattering) effects, 
Eqs.~\eqref{eqn:Tab} and \eqref{eqn:Tag} constitute a unified 
approach to the problem of parametrizing the hadronic 
scattering and photoproduction amplitudes. 

We pause here to make several remarks about the analytic form 
of the parametrization and its use in the present study. We 
first note that since the CM~K-matrix $\Kbar_{\sigma\gamma}(W)$ 
is a polynomial in the center-of-mass energy, $W$, an entire 
function, non-analytic points in the complex-$W$ plane are all 
a result of the hadronic rescattering matrix, $[1-\Kbar C]^{-1}$. 
This matrix has branch points and poles consistent with two-body 
and quasi-two-body unitarity~\cite{Eden:1966sm-1}. The 
quasi-two-body channels, \pD\ and \rn\ model the three-body \ppn\ 
channel only approximately. We use Eqs.~\eqref{eqn:Tab} and 
\eqref{eqn:Tag} as follows.  The parameters of the hadronic 
CM~K-matrix, $\Kbar_{\alpha,\beta}$, where $\alpha$ and $\beta$ may 
include (depending on the partial wave) $\pn,\pD,\rn$, and $\en$ 
are fixed by fitting the $\pntpn$ and \pnten\ data as in 
Ref.~\cite{Arndt:2006bf}. The $\pi$-photoproduction data is then 
fitted~\cite{limits} by varying only the parameters of the electromagnetic 
CM~K-matrix elements, $\Kbar_{\sigma\gamma}$, where $\sigma$ 
includes the channels $\pn,\pD,\rn$, and $\en$.

This approach differs markedly from that adopted in 
Refs.~\cite{said2002,Dugger:2007bt,Dugger:2009pn,Workman:2011vb}. 
There, the fit form, motivated by a multichannel Heitler K-matrix 
approach~\cite{Workman:2005eu},
\begin{equation}
        M \; = \; ({\rm Born} + A)(1 + i T_{\pi N} ) + B T_{\pi N} ,
\label{eq:q1}
\end{equation}
was modified to include a term 
\begin{equation}
        (C + iD)( {\rm Im} T_{\pi N} - |T_{\pi N} |^2 ), 
\label{eq:q2}
\end{equation}
where $T_{\pi N}$ is the elastic $\pi N$ scattering partial-wave 
amplitude associated with the pion-photoproduction multipole 
amplitude $M$. The added piece, which grows with the $\pi N$ reaction
cross section, was found to improve the fit at energies above the 
$\eta N$ threshold. Each of the phenomenological terms $A$ to $D$ was
parameterized as a polynomial in energy, having the correct threshold
behavior.

In the new form, terms $A$ and $B$ have been effectively replaced by
a single CM~K-matrix element, $\Kbar_{\pi\gamma}$. Conversely, the
influence of channels opening above $\pi N$ is now (more correctly)
associated with individual channels ($\eta N$, $\pi \Delta$, and $\rho 
N$) rather than a single term. Large cancellations found to occur 
between the Born, $A$, and $B$ terms suggested that a more economical 
parametrization is possible~\cite{Workman:2005eu}. In fact, the
CM~K-matrix form provides a better overall fit to the data with
fewer free parameters, as we will show in the next section.

%%%%%%%%%%%%%%%%%%%%%%%%%%%%%%%%%%%%%%%%%%%%%%%%%%%%%%%%%%%%%%%%%%
\section{Fit results and multipole amplitudes}
\label{sec:mpole}

In Table~\ref{tab:tbl1}, the fit quality and number of searched parameters 
is compared for the two fit forms discussed in the previous section. Here 
we have used the same database to 2.7~GeV as was used in generating 
solution SN11~\cite{Workman:2011vb}. The present CM12 form requires fewer 
parameters to achieve a slightly better data fit. The energy-dependence of 
this result was tested by repeating the analyses over three different 
energy ranges. 

As the fit employs polynomial functions for $A$ to $D$, of SN11, or the 
electromagnetic CM~K-matrix elements of CM12, a subjective criteria is 
required to determine the order of polynomials fitted.  In the fits to 
2.7~GeV, the order of polynomial functions was increased until further 
additions improved the overall $\chi^2$ by 50 or less. This same criteria 
was used in both the SN11 and CM12 fits, in order to have a basis for 
comparison. As more parameters were searched, their ability to improve 
the fit diminished. In the fits to lower energy limits, parameters were 
removed in steps, again with the condition that removing a parameter 
should not increase the overall $\chi^2$ by more than 50.

\begin{figure*}[th]
\centerline{
\includegraphics[height=0.42\textwidth, angle=90]{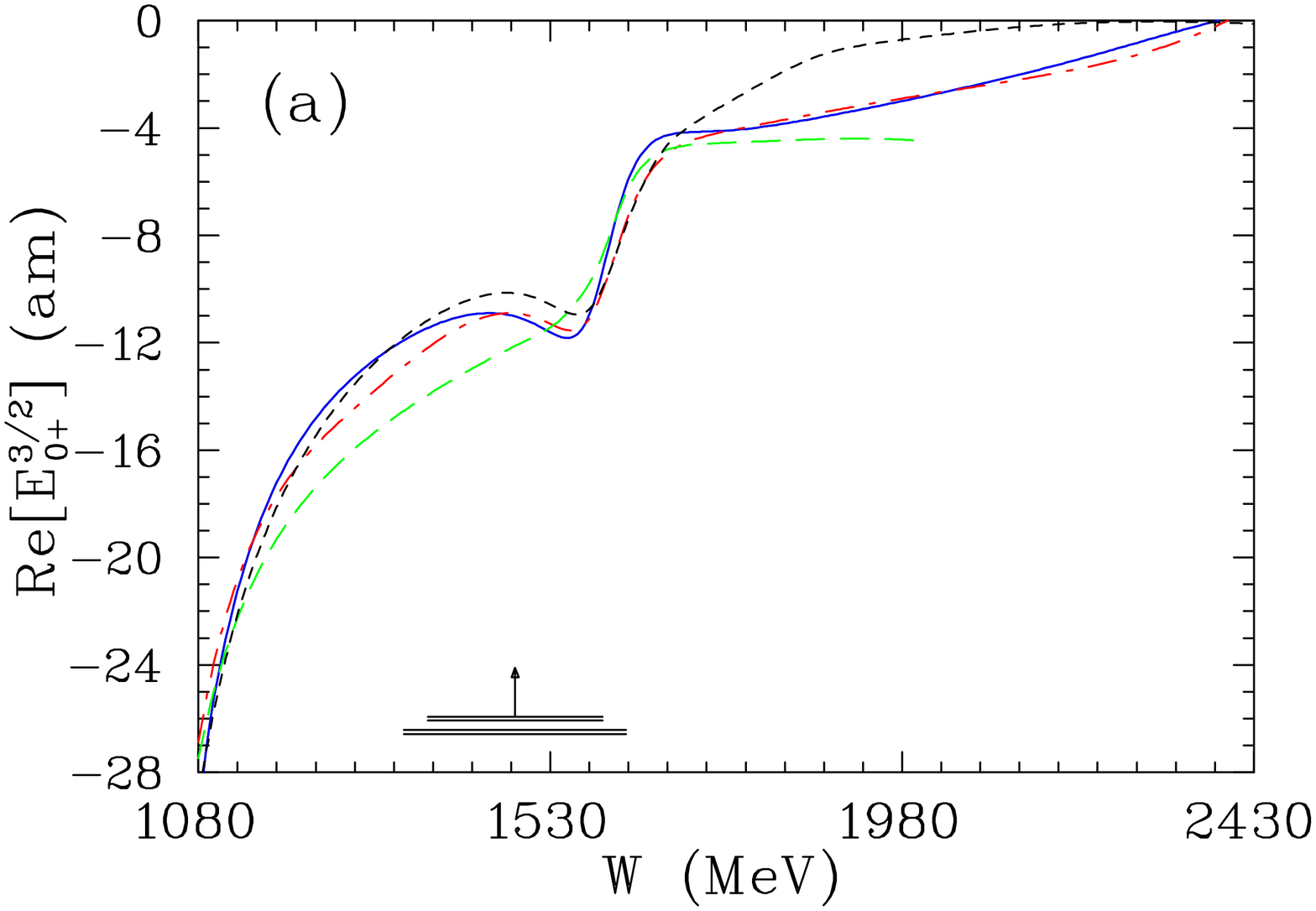}\hfill
\includegraphics[height=0.42\textwidth, angle=90]{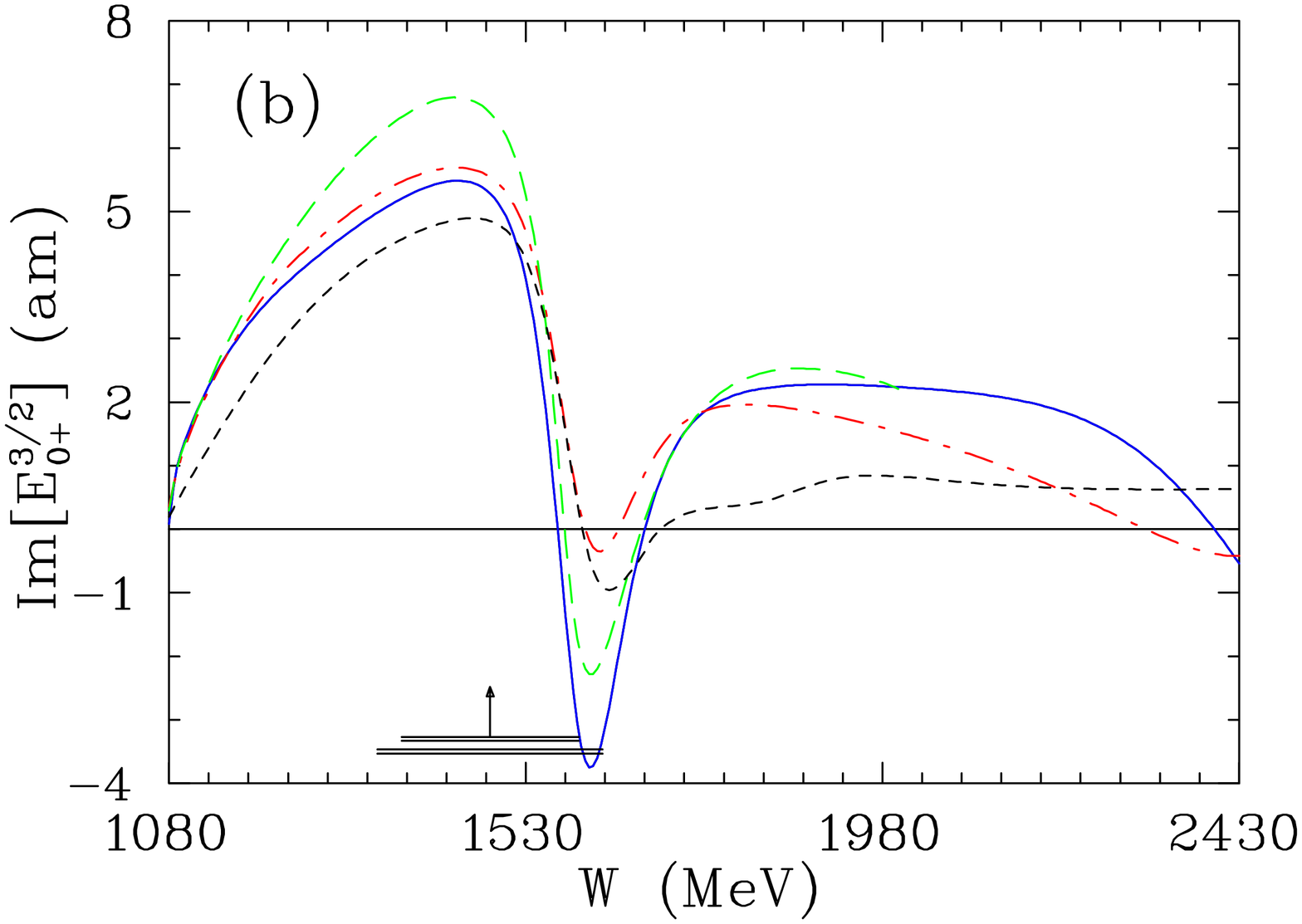}}
\centerline{
\includegraphics[height=0.42\textwidth, angle=90]{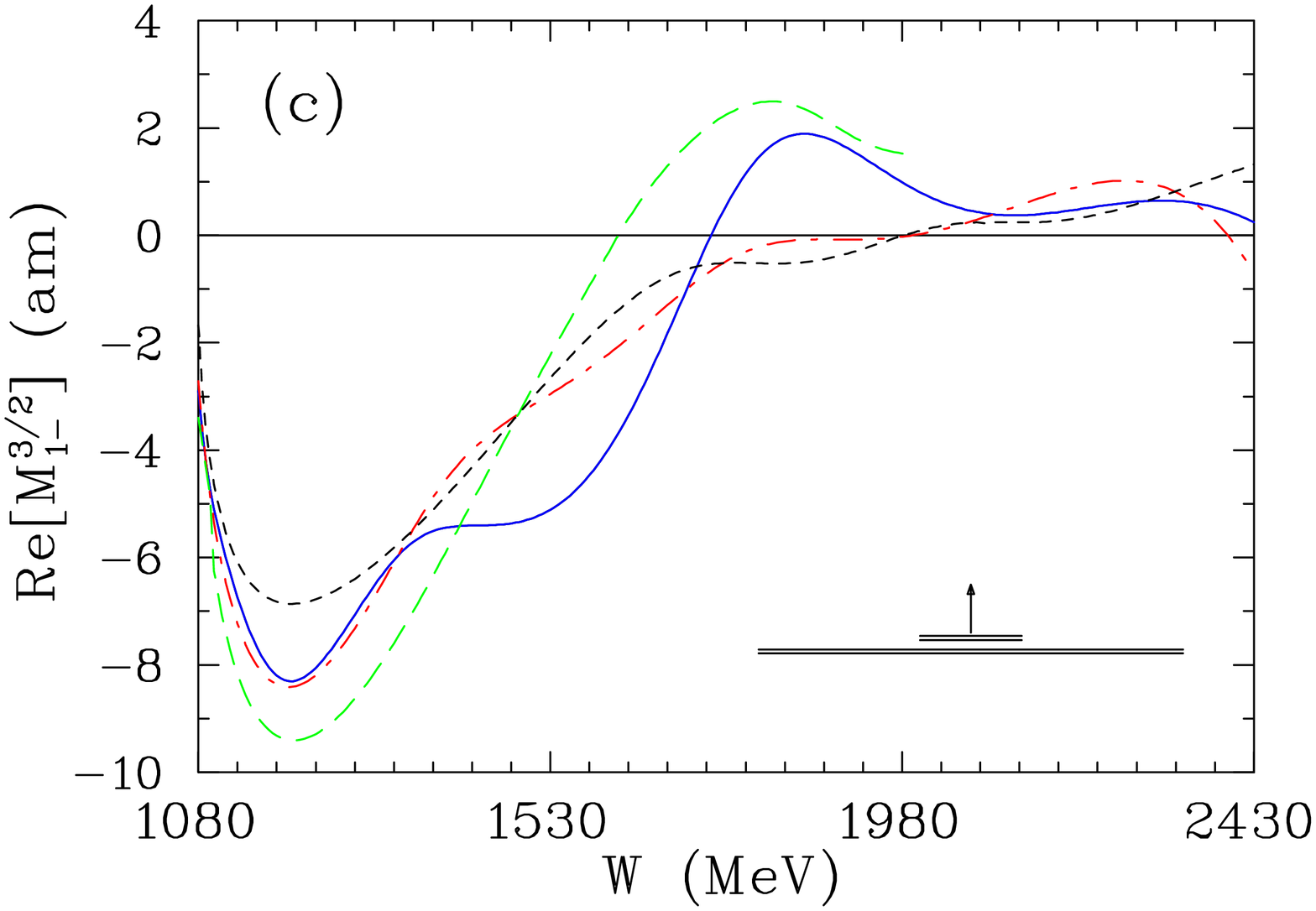}\hfill
\includegraphics[height=0.42\textwidth, angle=90]{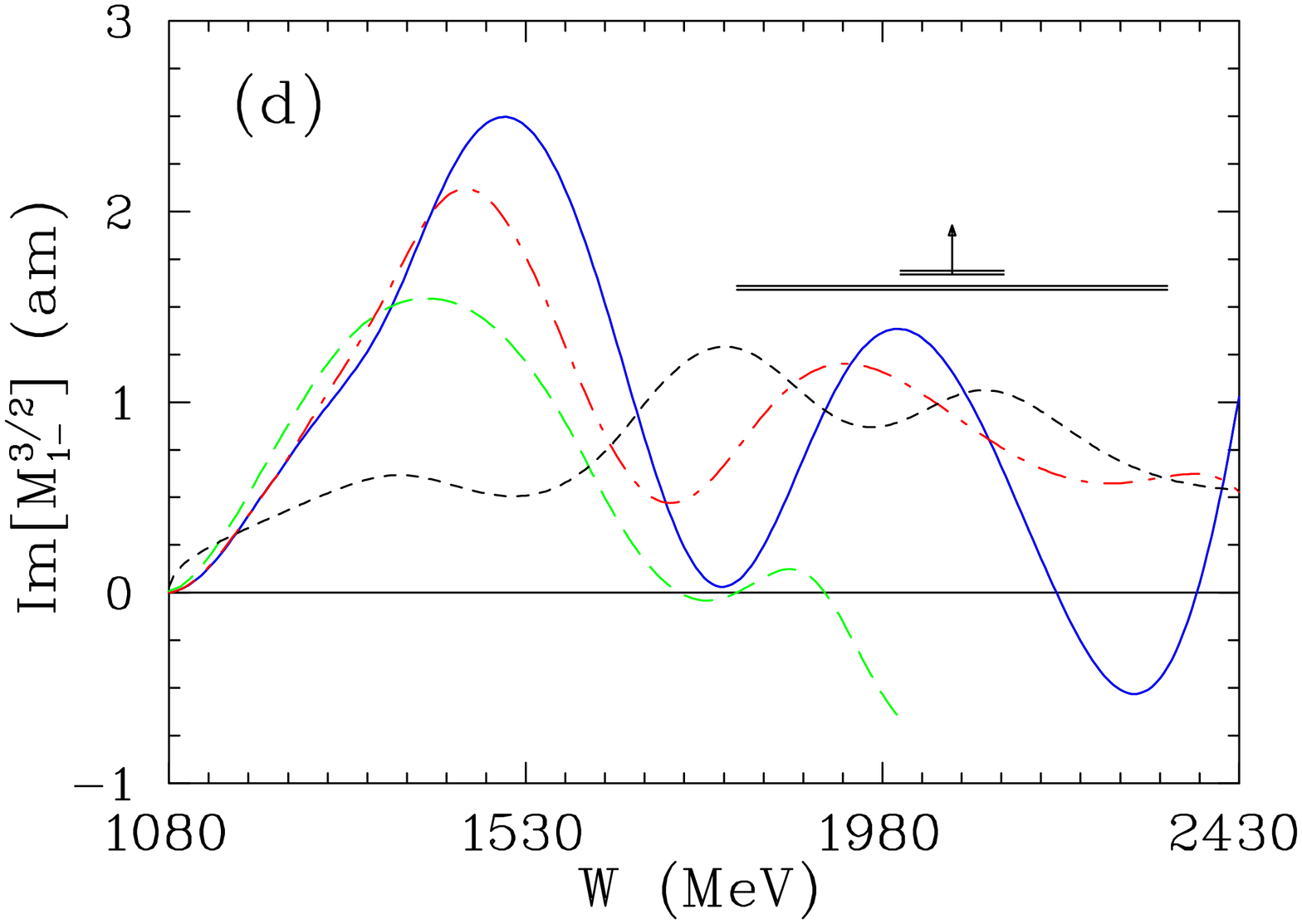}}
\centerline{
\includegraphics[height=0.42\textwidth, angle=90]{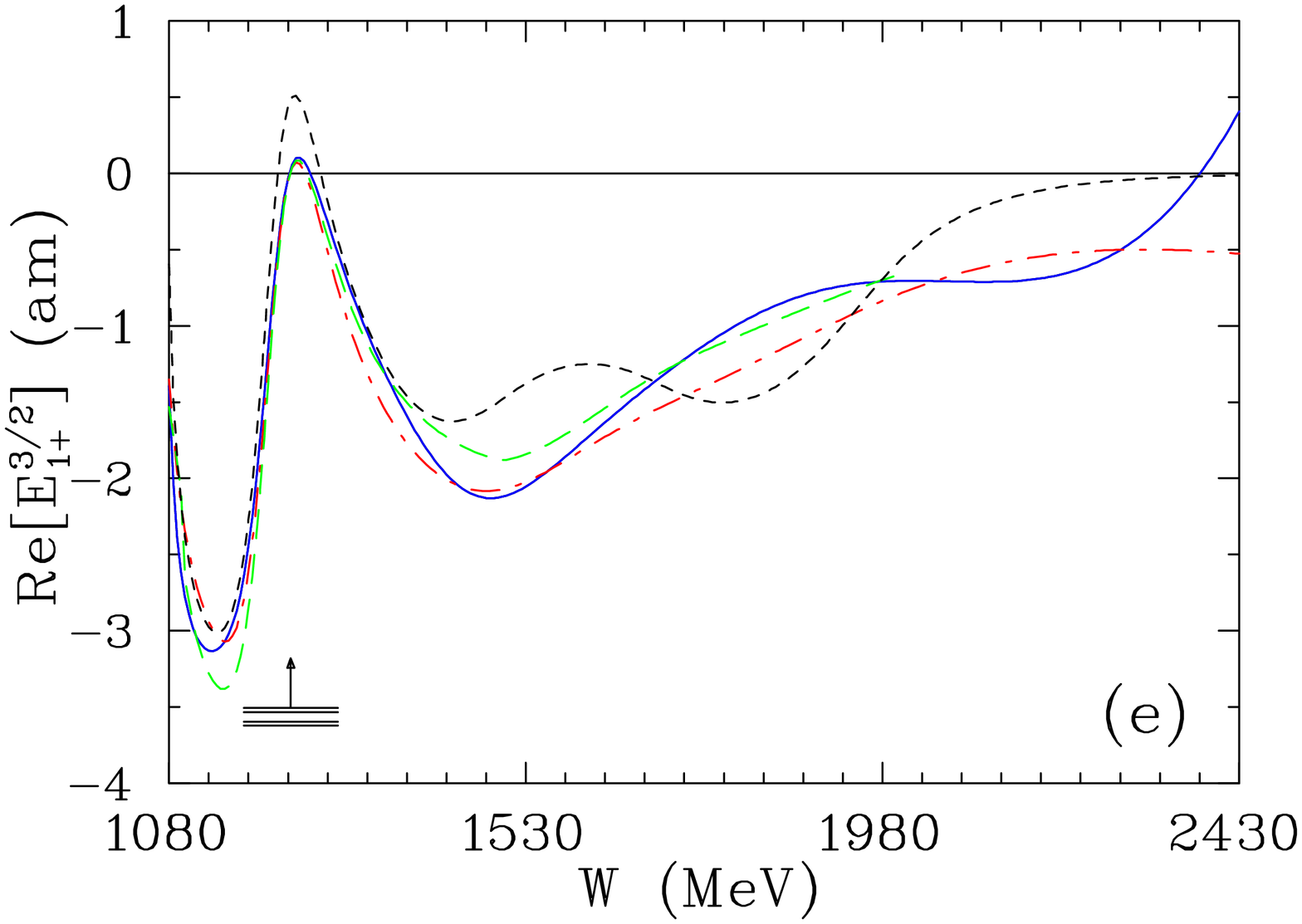}\hfill
\includegraphics[height=0.42\textwidth, angle=90]{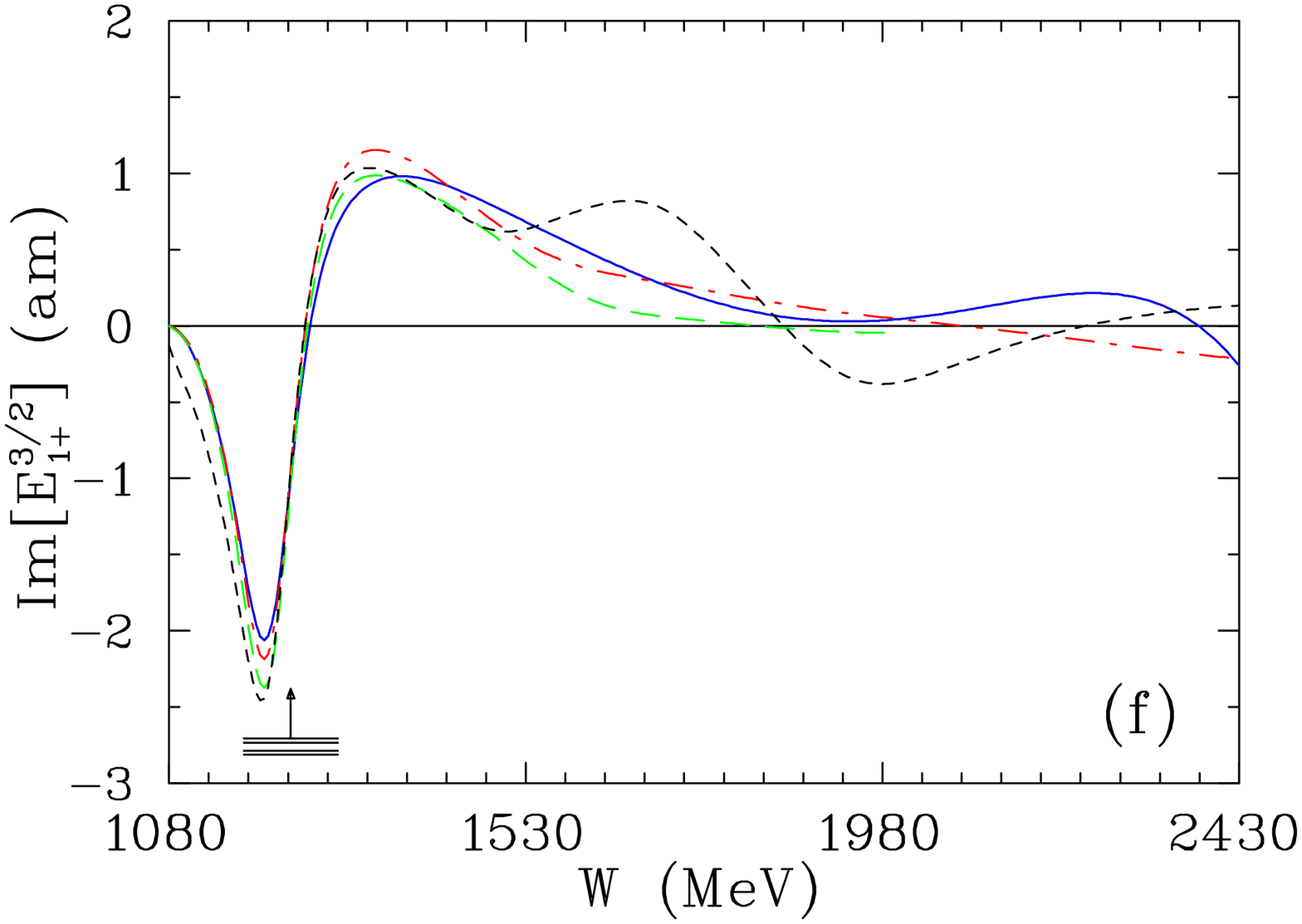}}
\centerline{
\includegraphics[height=0.42\textwidth, angle=90]{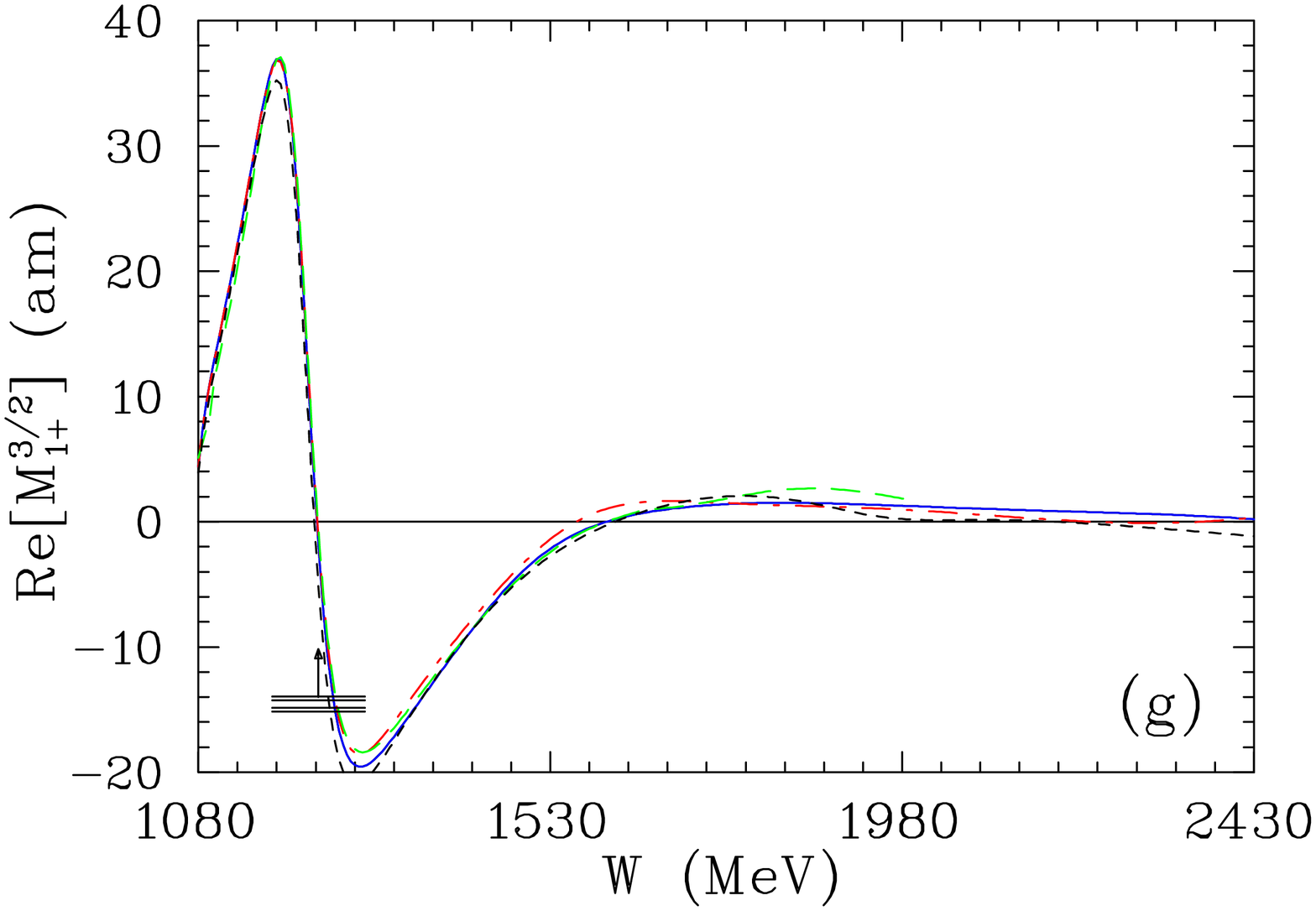}\hfill
\includegraphics[height=0.42\textwidth, angle=90]{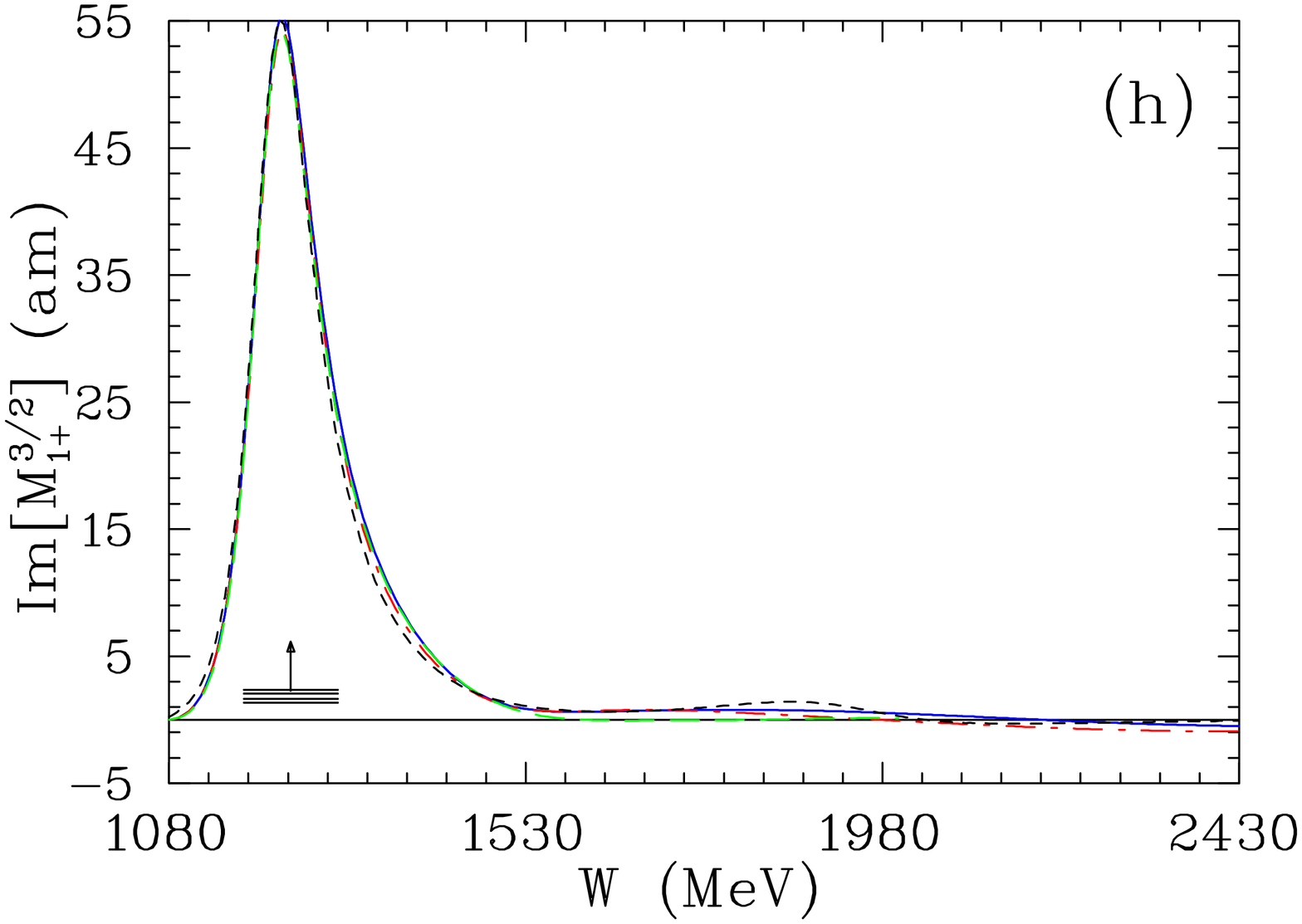}}
\caption{(Color online) I=3/2 multipole amplitudes from 
        threshold to $W$ = 2.43~GeV ($E_{\gamma}$ = 2.7~GeV). 
        Solid (dash-dotted) lines correspond to the SN11 (CM12) 
        solution.  Short-dashed (dashed) lines give BG2010-02 
        solution~\protect\cite{BoGa} (MAID07~\protect\cite{maid}, 
        which terminates at $W$=2~GeV).  Vertical arrows 
        indicate resonance energies, $W_R$, and horizontal bars
        show full ($\Gamma$) and partial ($\Gamma_{\pi N}$)
        widths associated with the SAID $\pi N$ solution
        SP06~\protect\cite{Arndt:2006bf}. \label{fig:f1}}
\end{figure*}

\begin{figure*}[th]
\centerline{
\includegraphics[height=0.42\textwidth, angle=90]{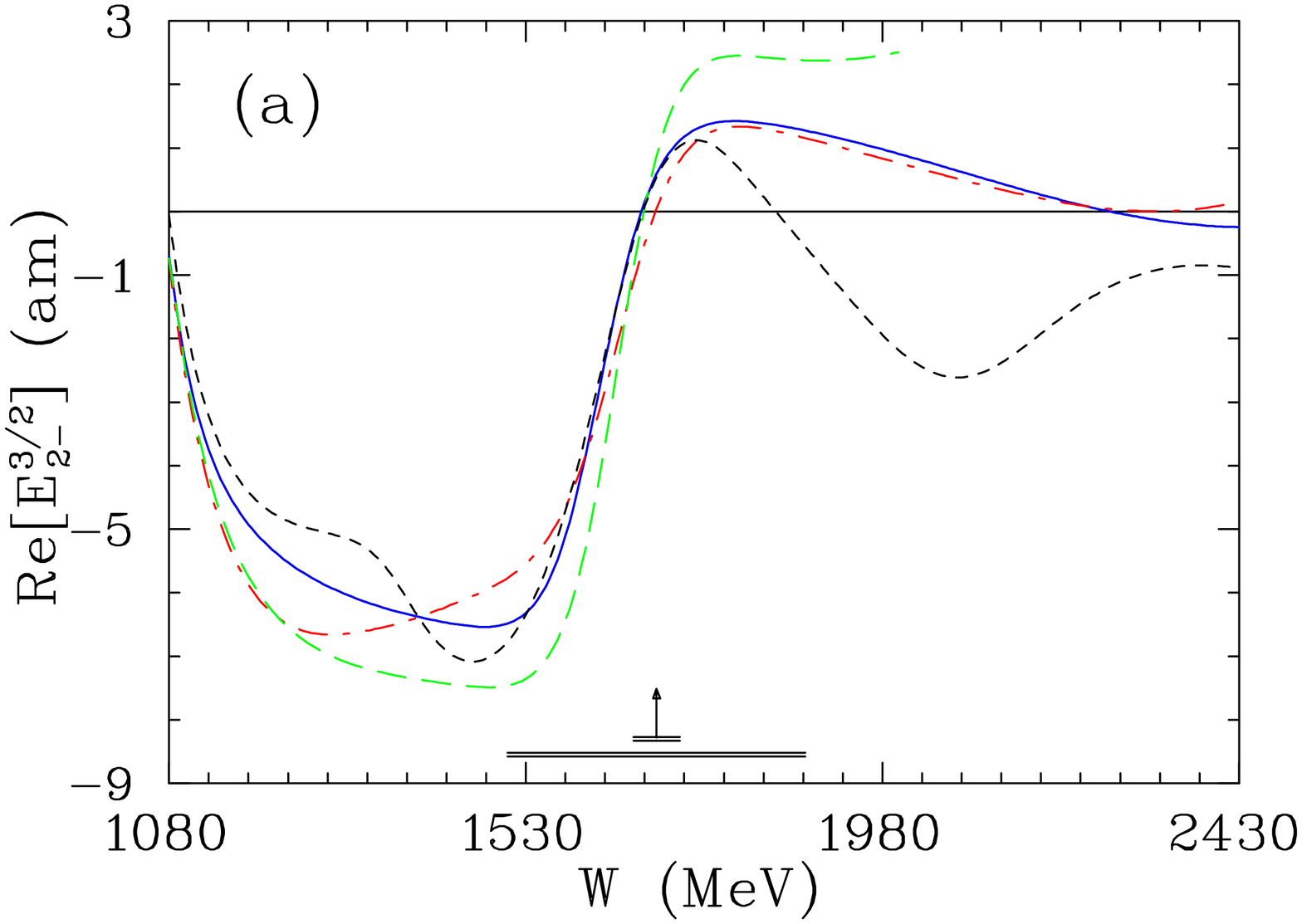}\hfill
\includegraphics[height=0.42\textwidth, angle=90]{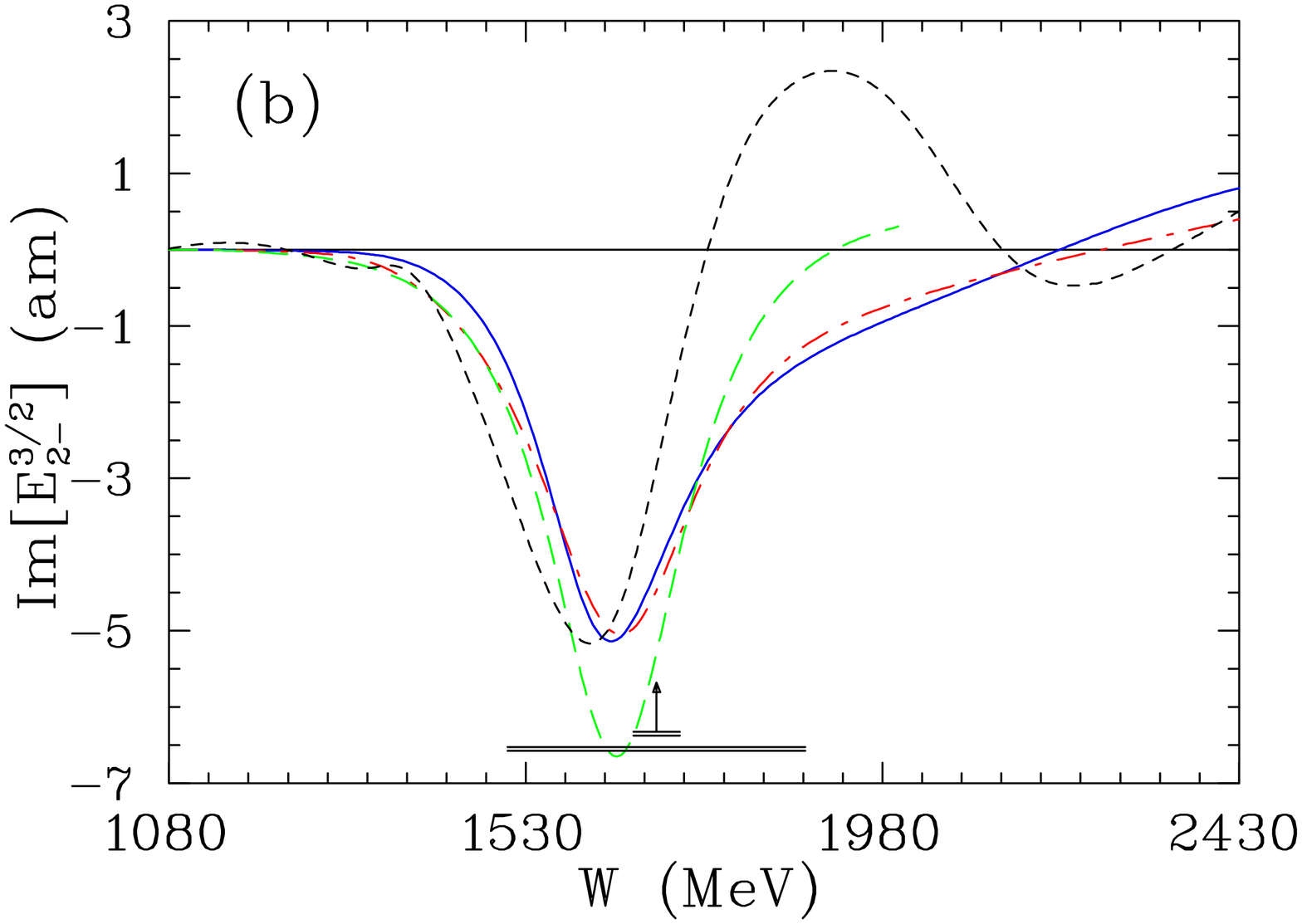}}
\centerline{
\includegraphics[height=0.42\textwidth, angle=90]{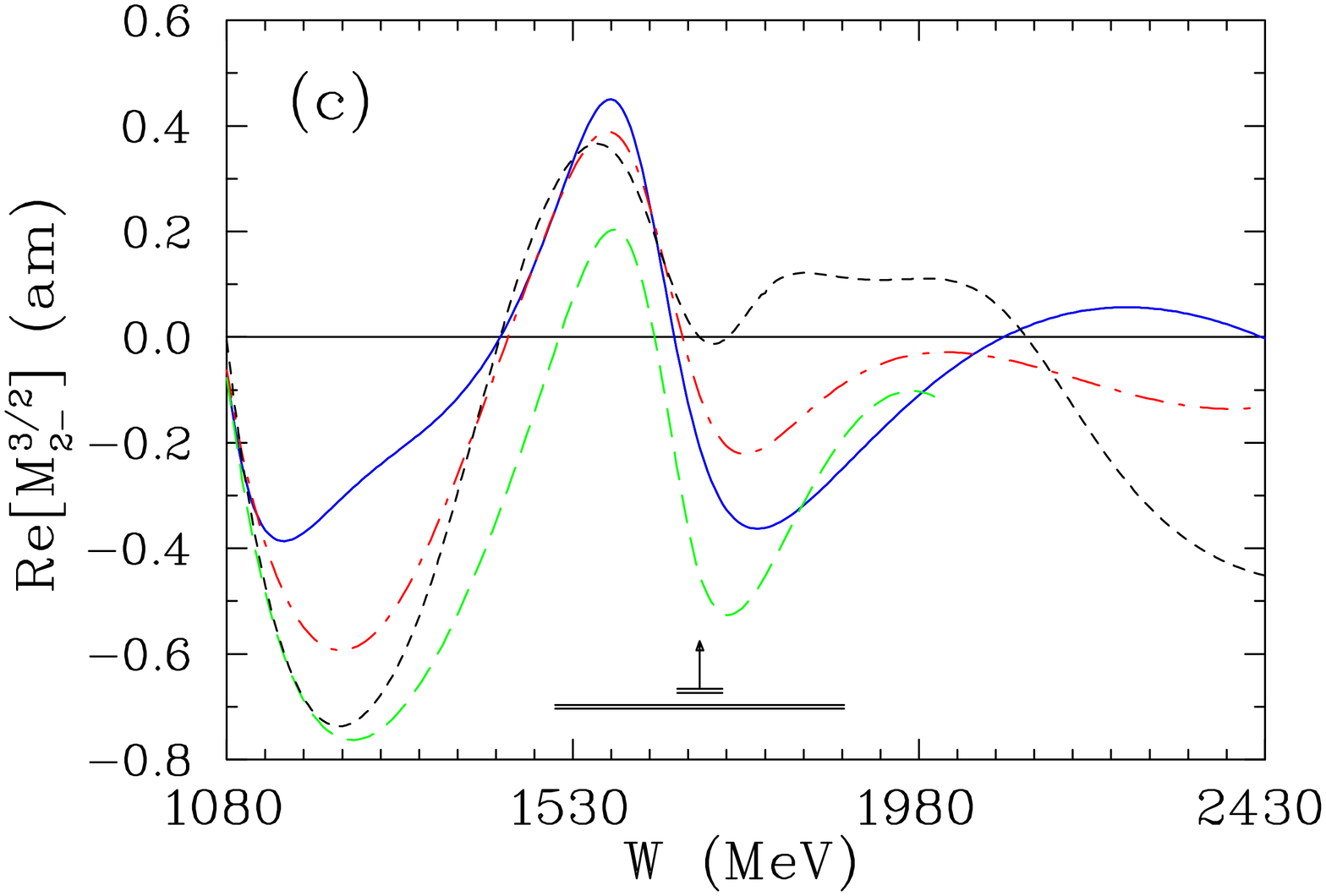}\hfill
\includegraphics[height=0.42\textwidth, angle=90]{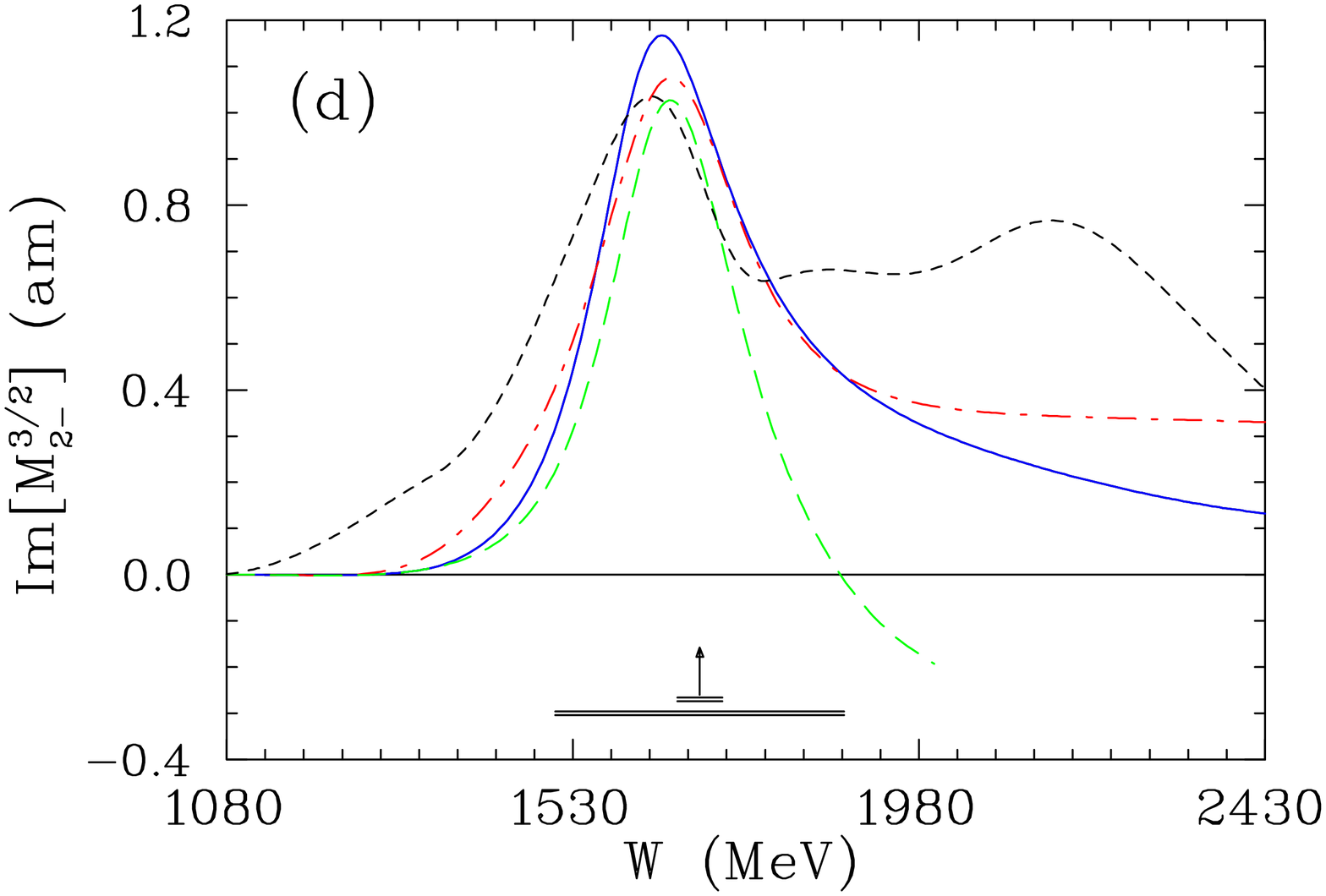}}
\centerline{
\includegraphics[height=0.42\textwidth, angle=90]{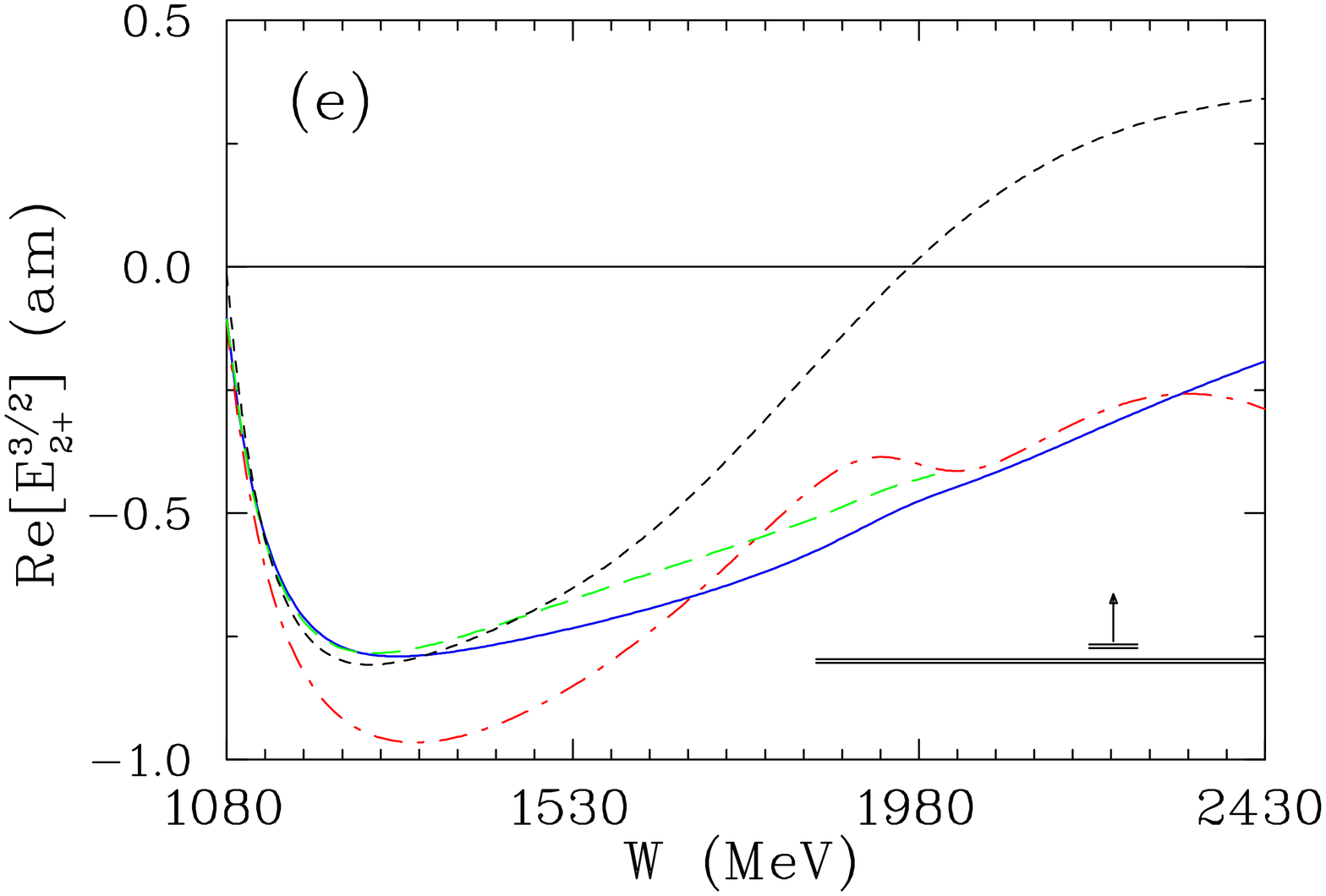}\hfill
\includegraphics[height=0.42\textwidth, angle=90]{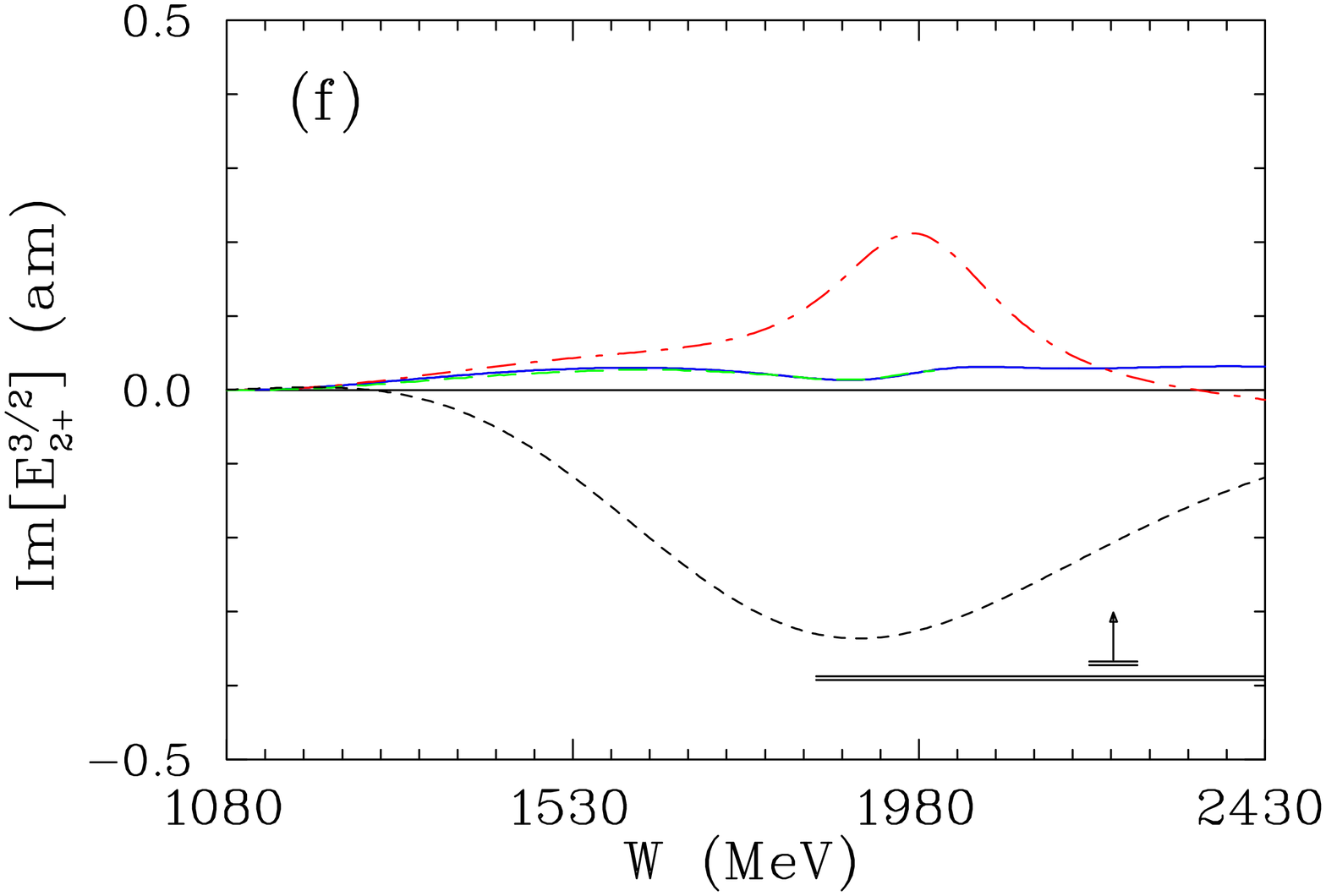}}
\centerline{
\includegraphics[height=0.42\textwidth, angle=90]{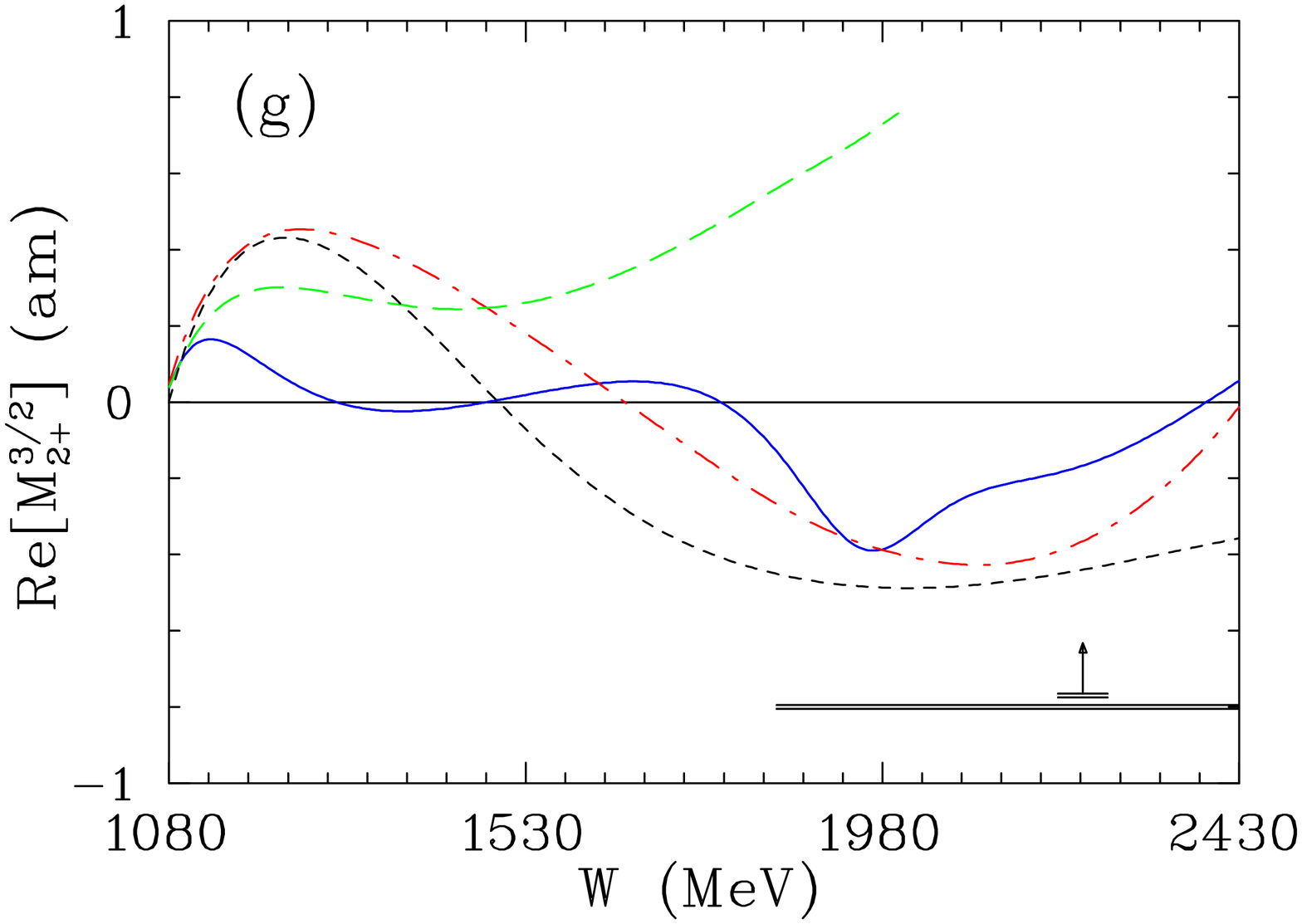}\hfill
\includegraphics[height=0.42\textwidth, angle=90]{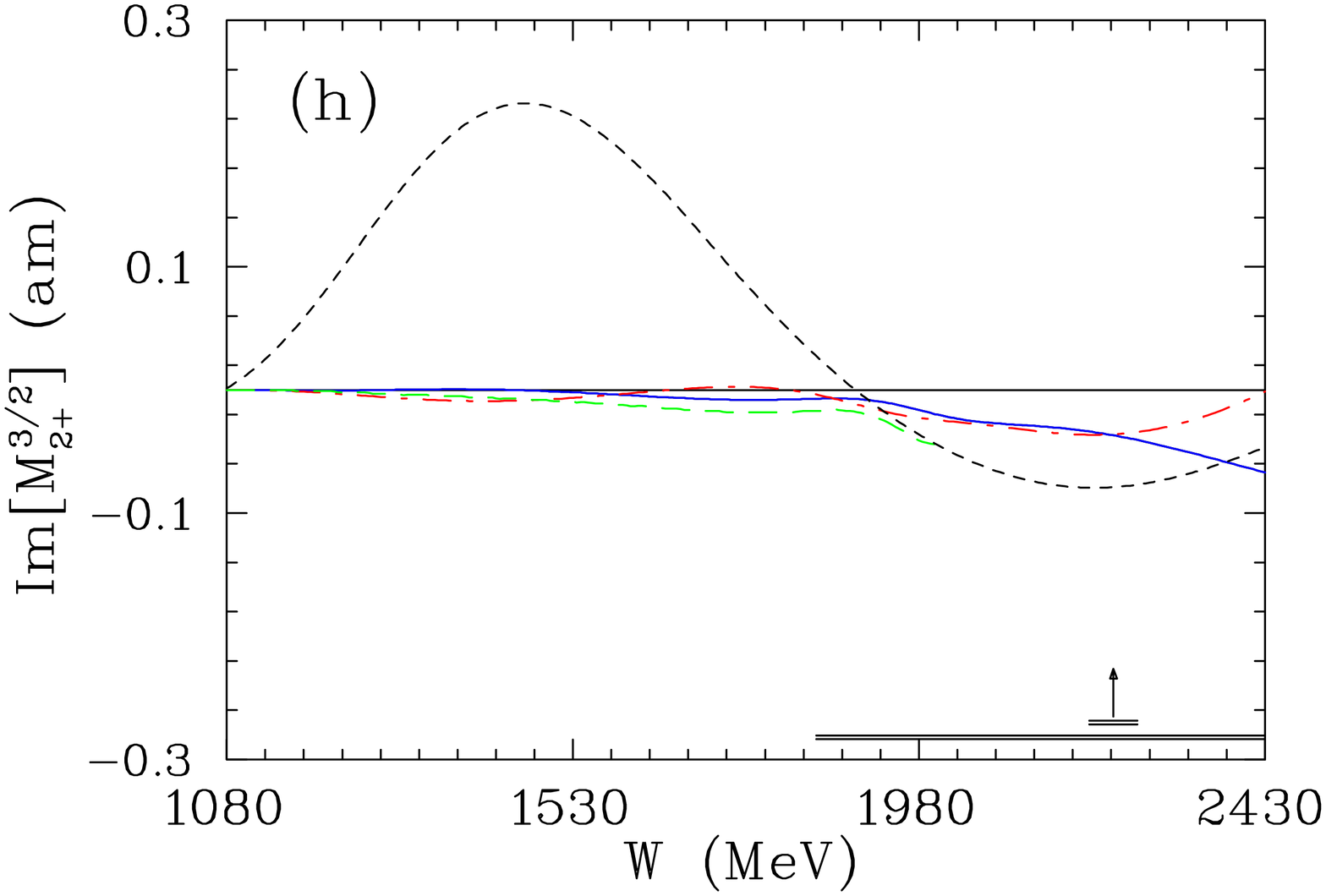}}
\caption{(Color online) Notation of the multipoles is the 
	same as in Fig.~\protect\ref{fig:f1}. \label{fig:f2}}
\end{figure*}

\begin{figure*}[th]
\centerline{
\includegraphics[height=0.42\textwidth, angle=90]{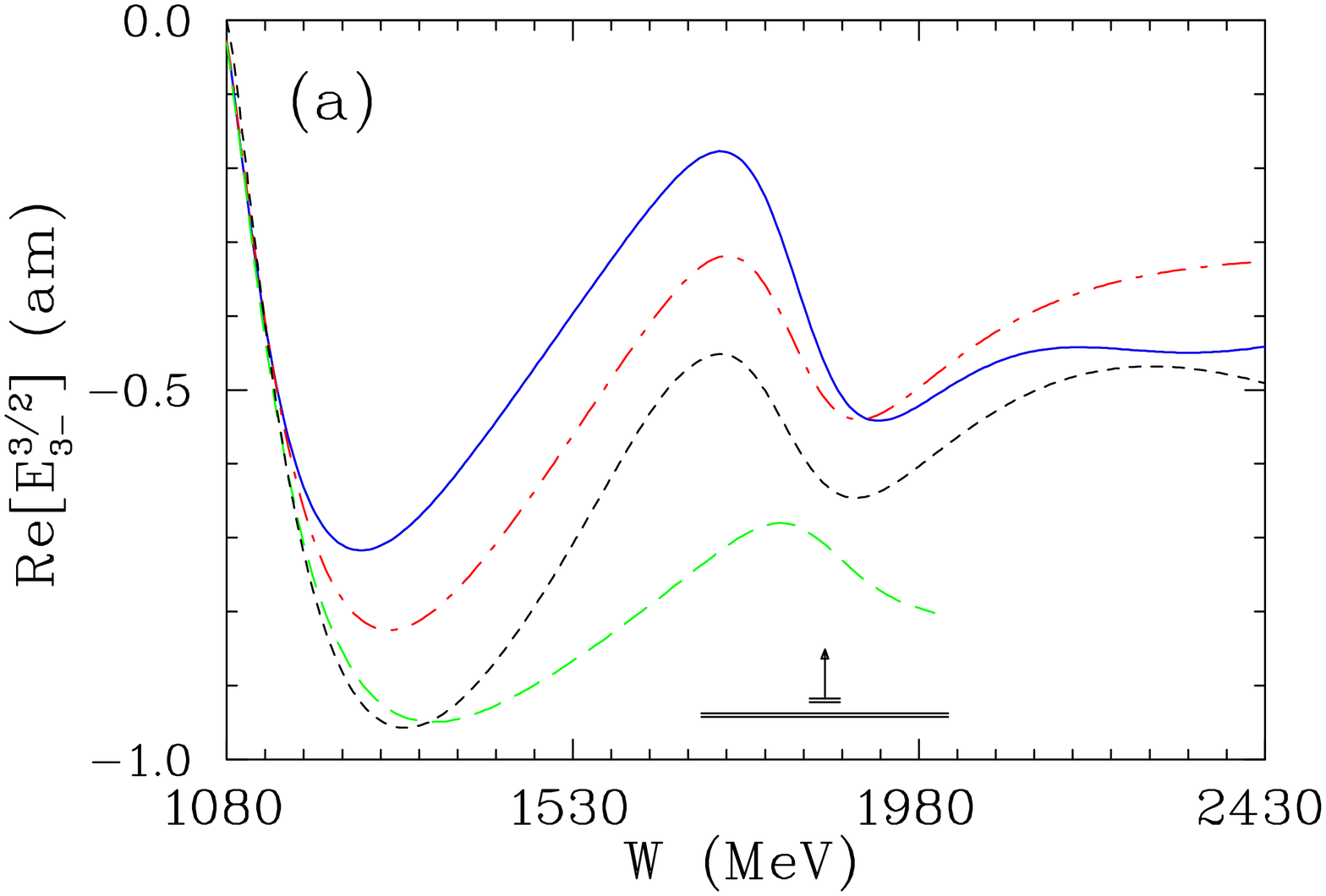}\hfill
\includegraphics[height=0.42\textwidth, angle=90]{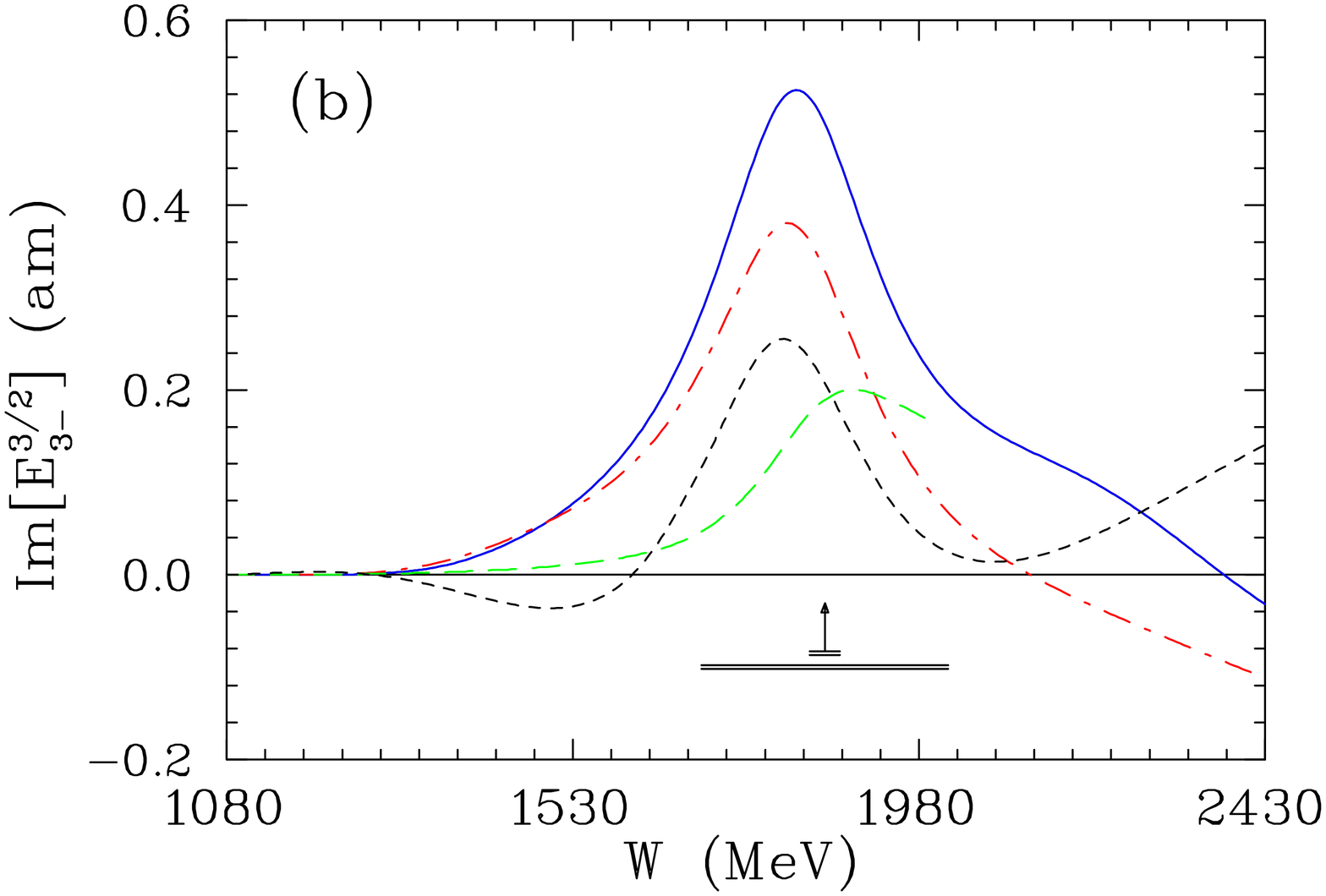}}
\centerline{
\includegraphics[height=0.42\textwidth, angle=90]{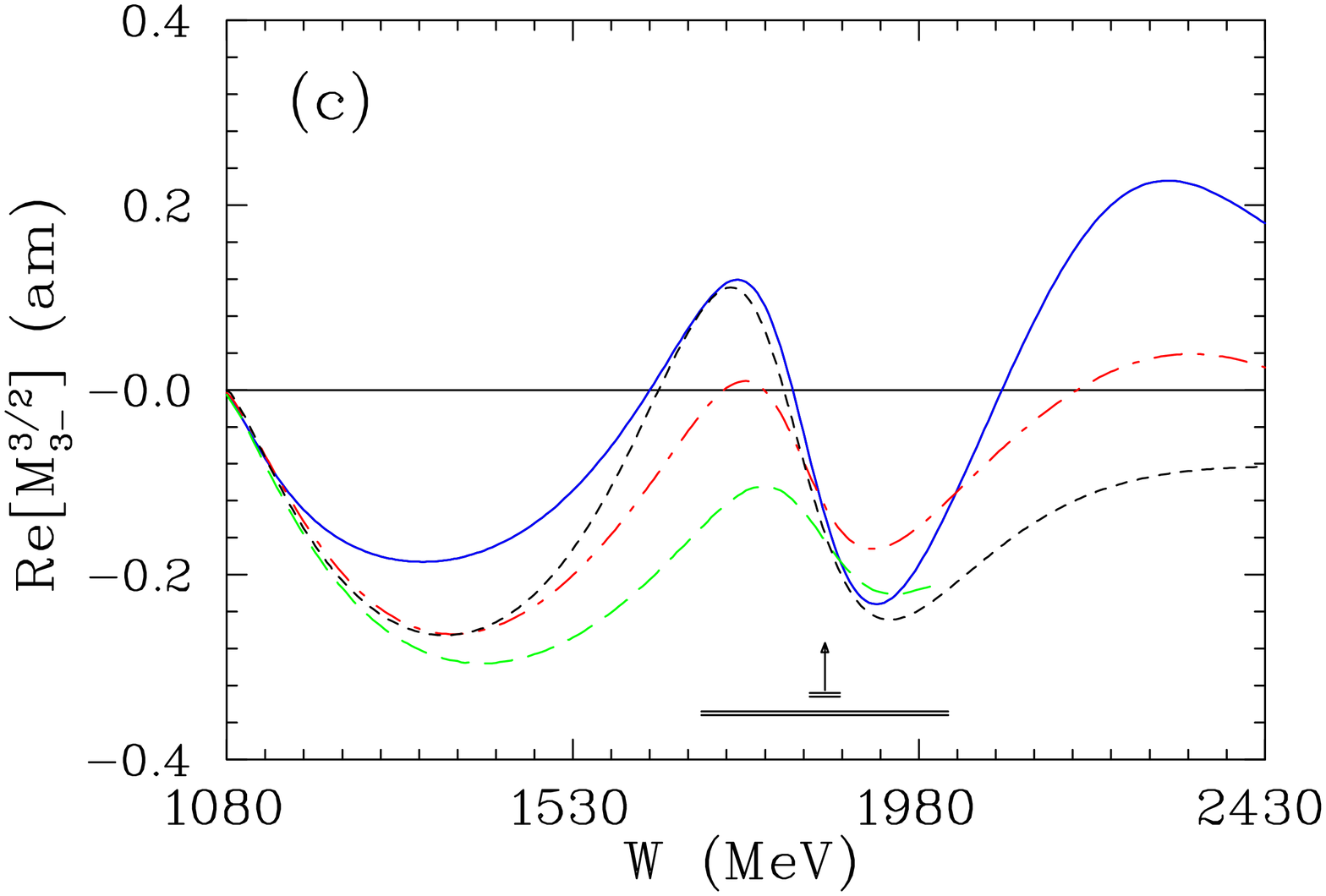}\hfill
\includegraphics[height=0.42\textwidth, angle=90]{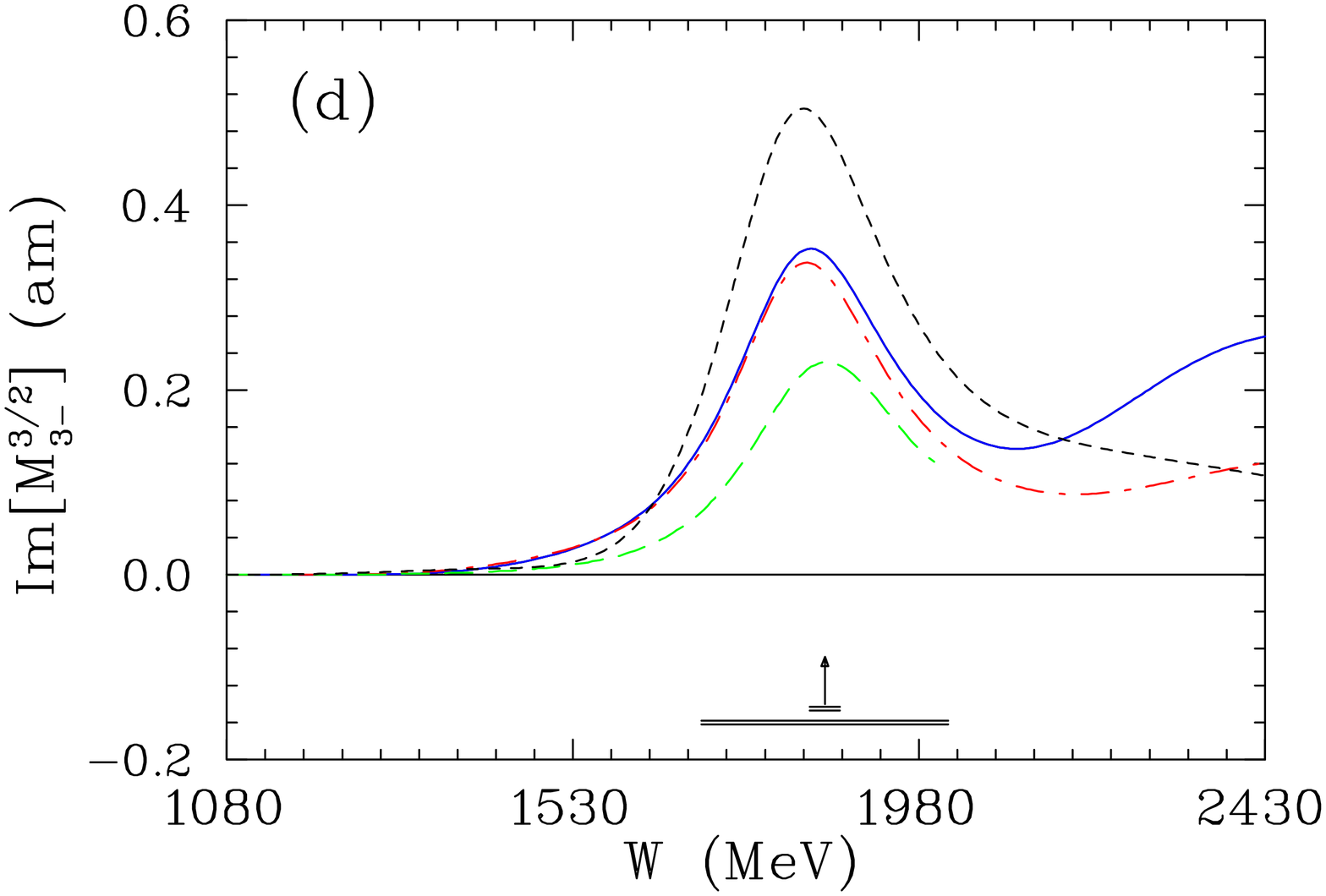}}
\centerline{
\includegraphics[height=0.42\textwidth, angle=90]{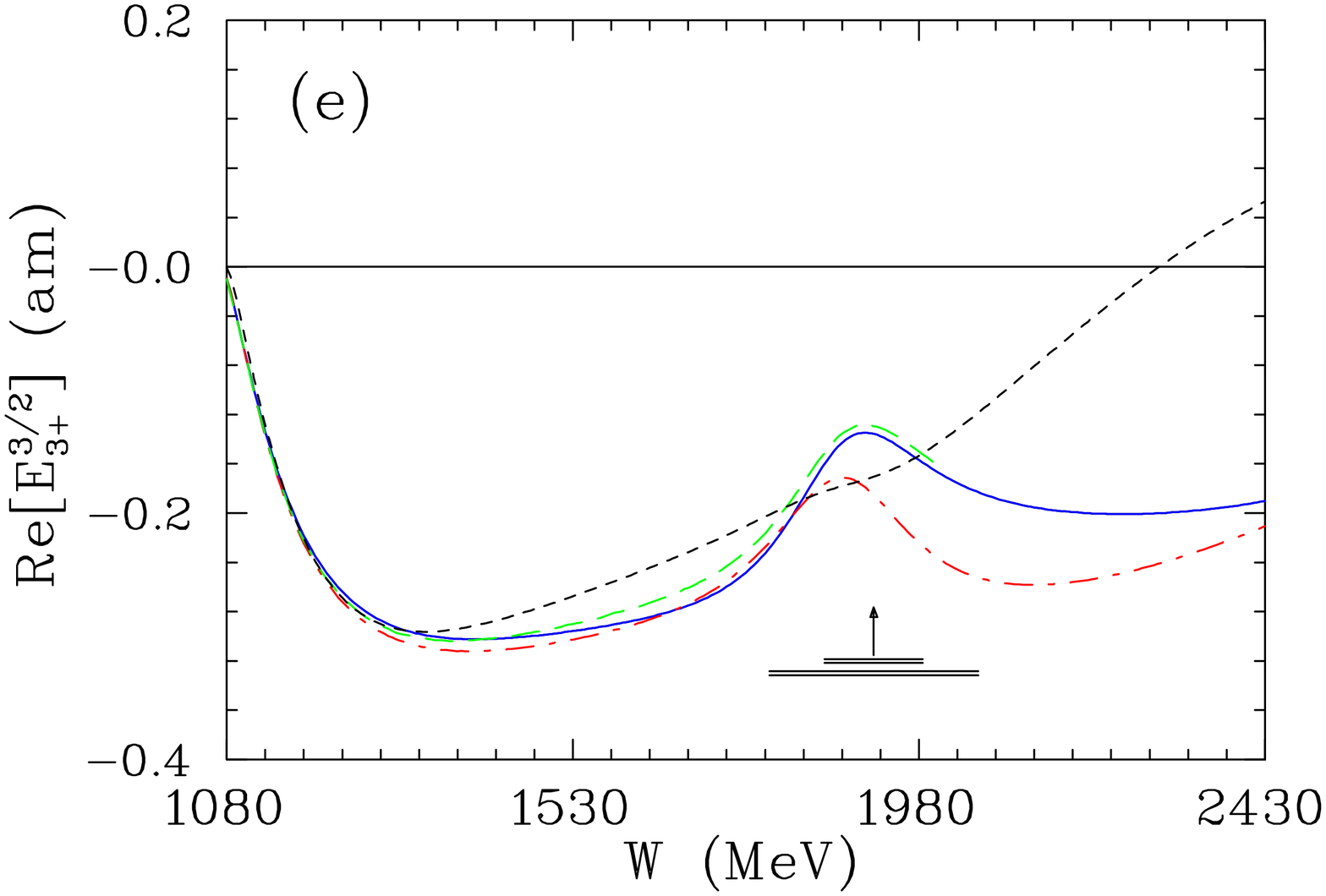}\hfill
\includegraphics[height=0.42\textwidth, angle=90]{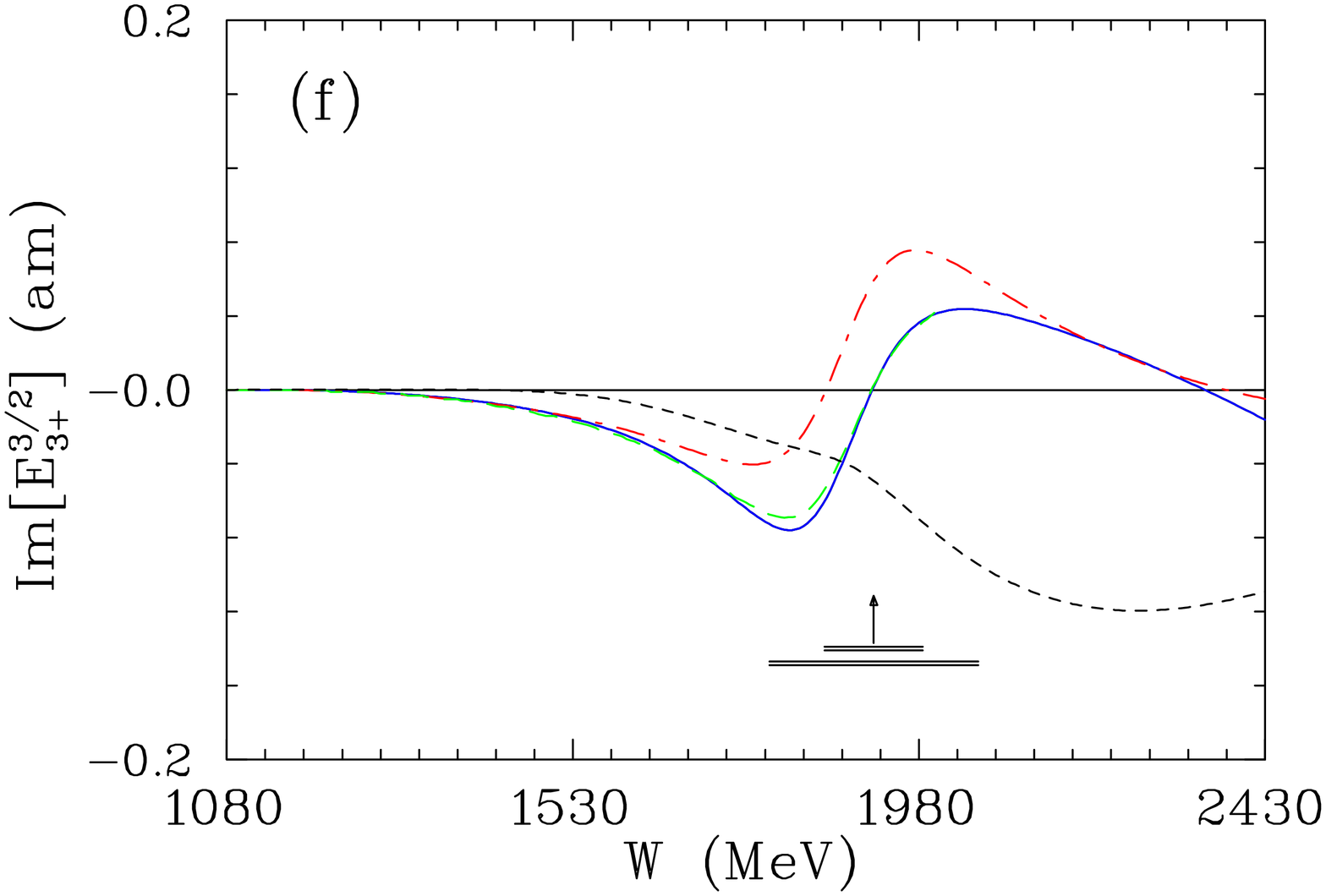}}
\centerline{
\includegraphics[height=0.42\textwidth, angle=90]{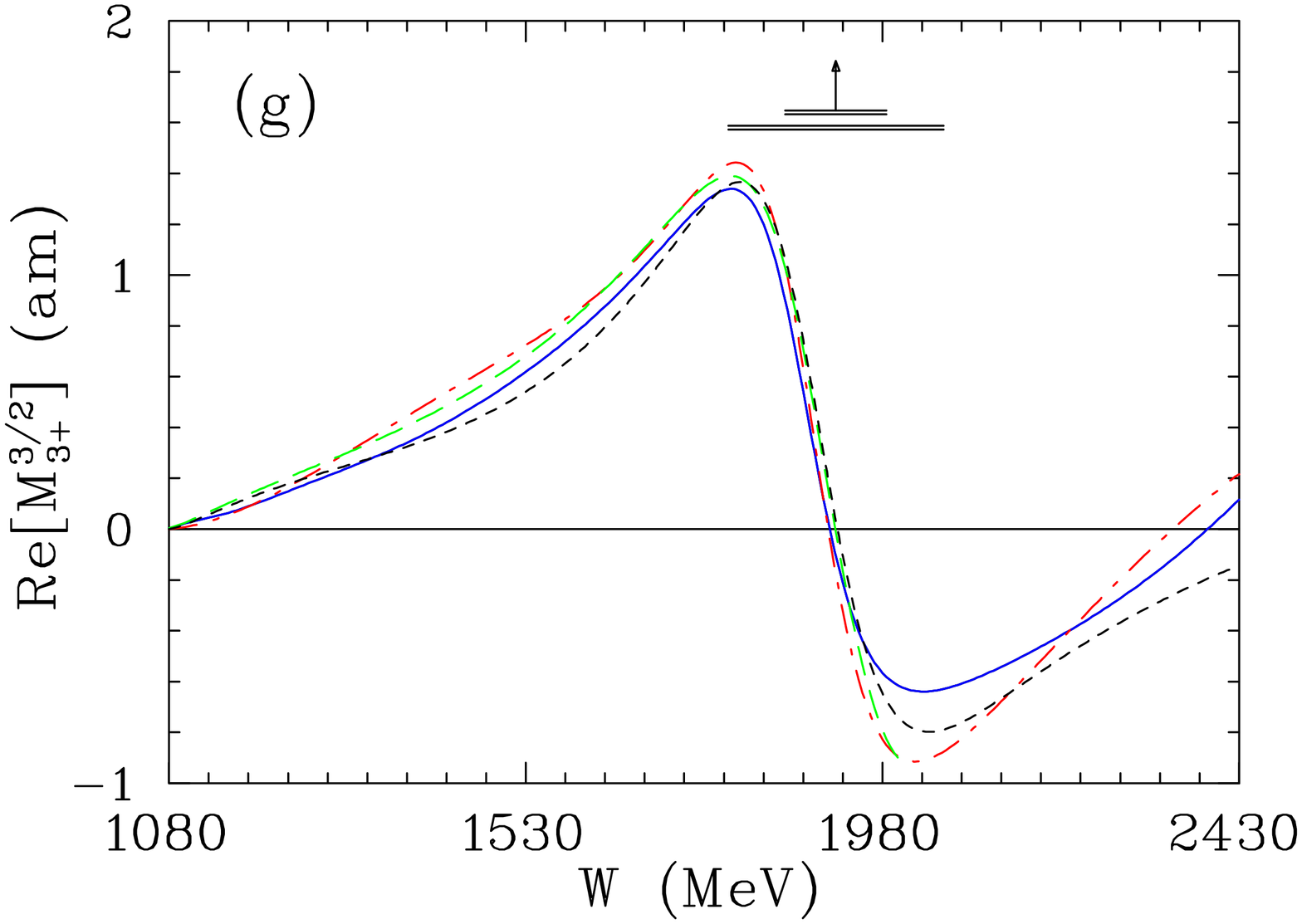}\hfill
\includegraphics[height=0.42\textwidth, angle=90]{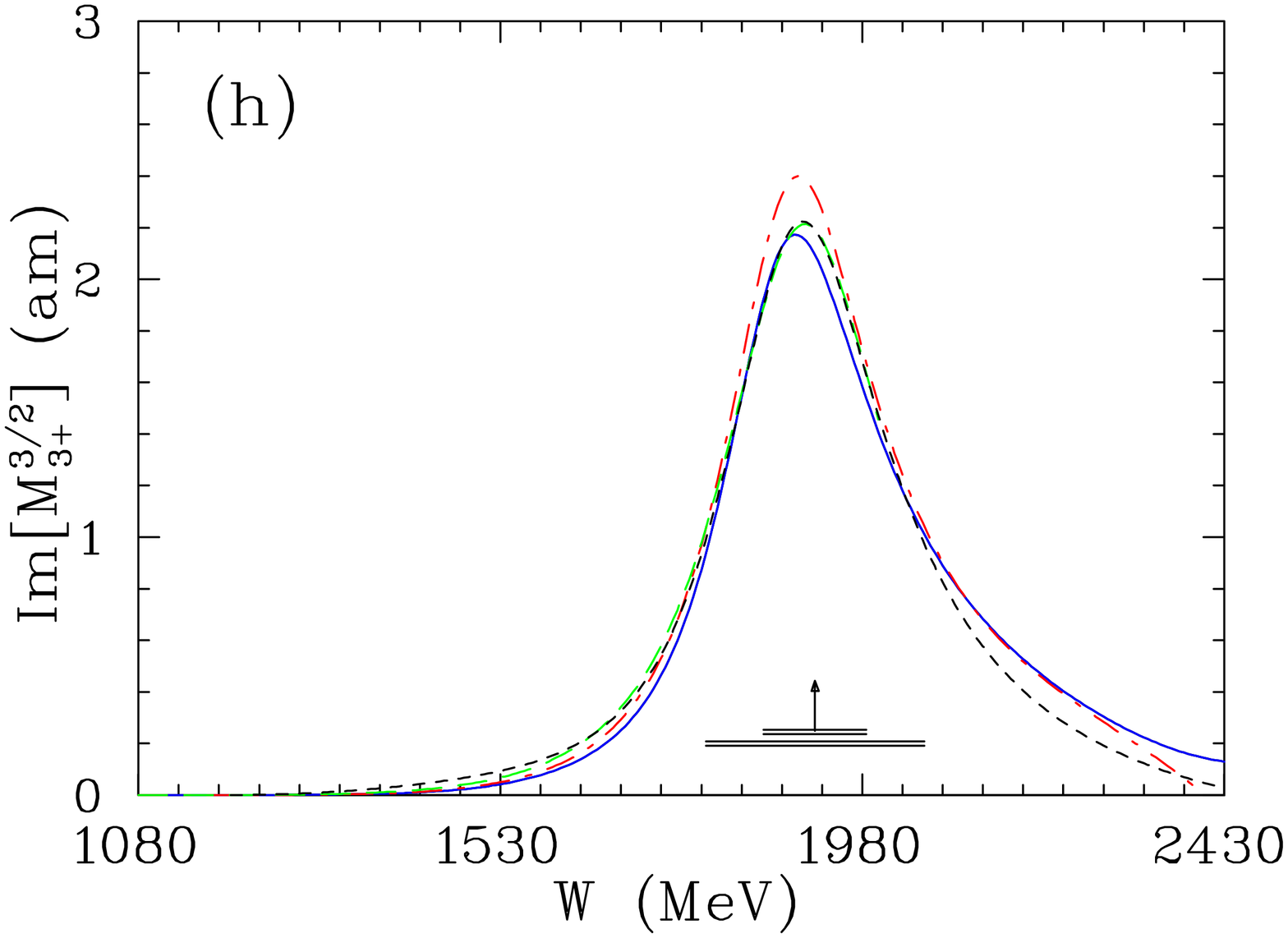}}
\caption{(Color online) Notation of the multipoles is the 
	same as in Fig.~\protect\ref{fig:f1}.\label{fig:f3}}
\end{figure*}

\begin{figure*}[th]
\centerline{
\includegraphics[height=0.42\textwidth, angle=90]{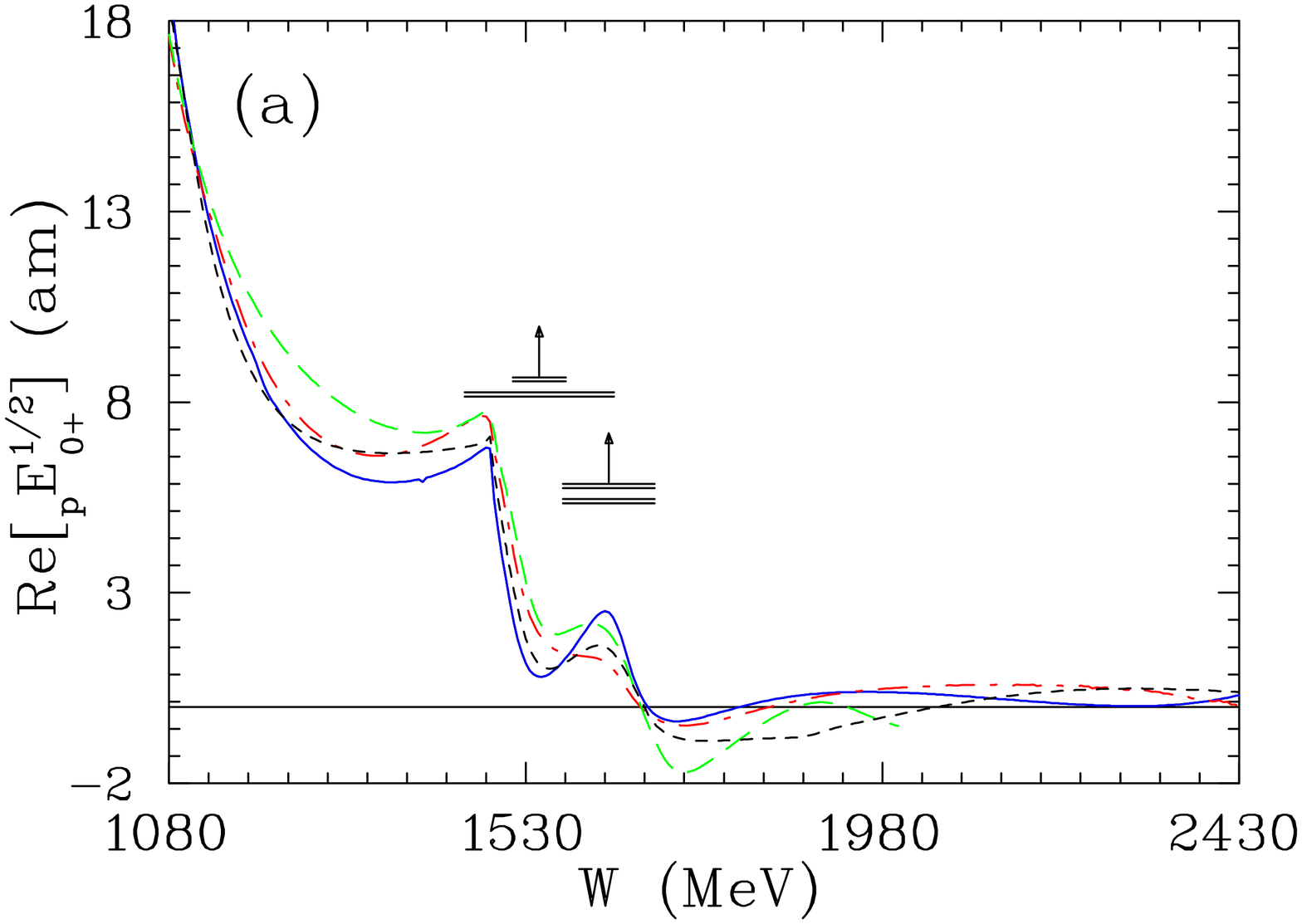}\hfill
\includegraphics[height=0.42\textwidth, angle=90]{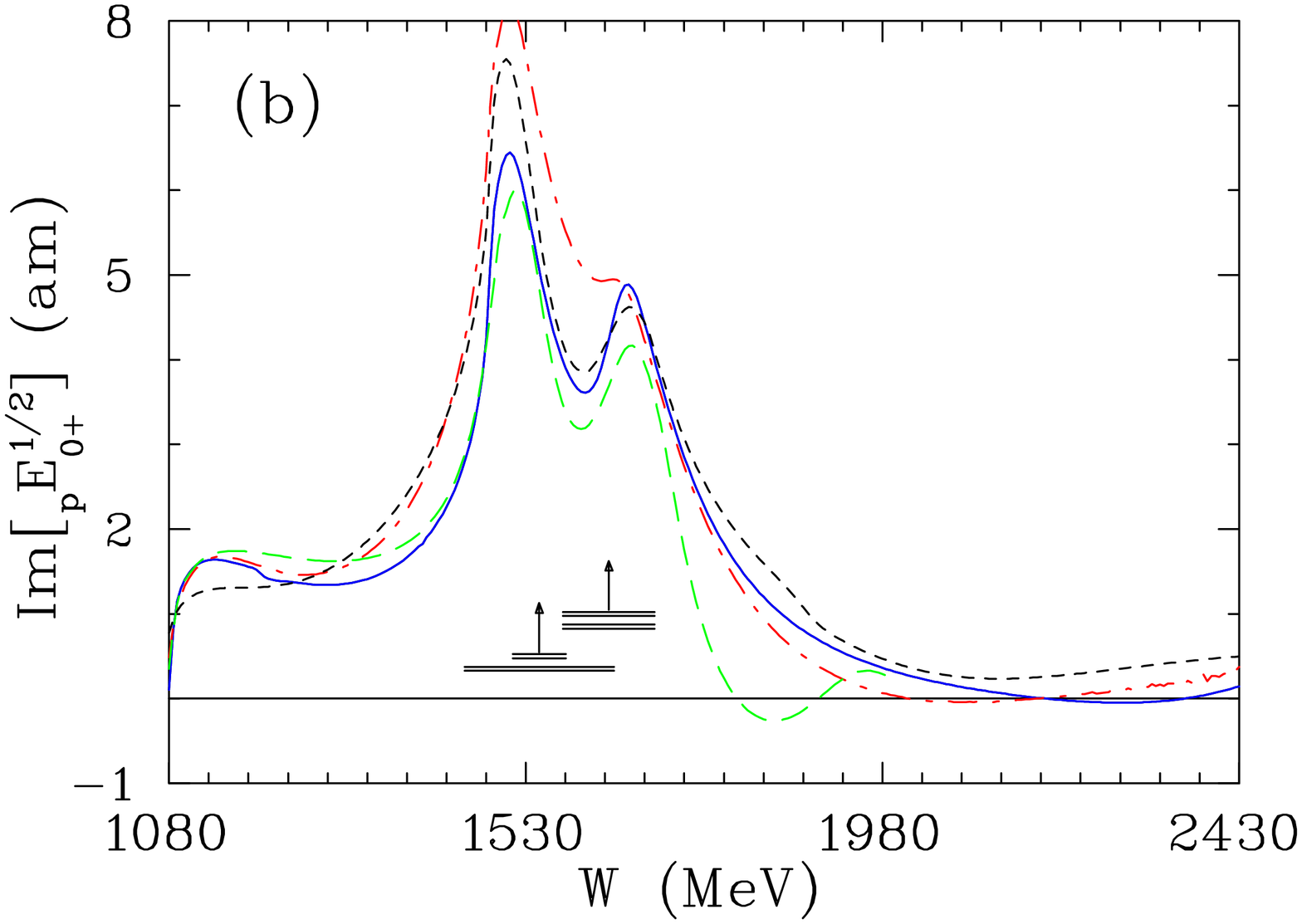}}
\centerline{
\includegraphics[height=0.42\textwidth, angle=90]{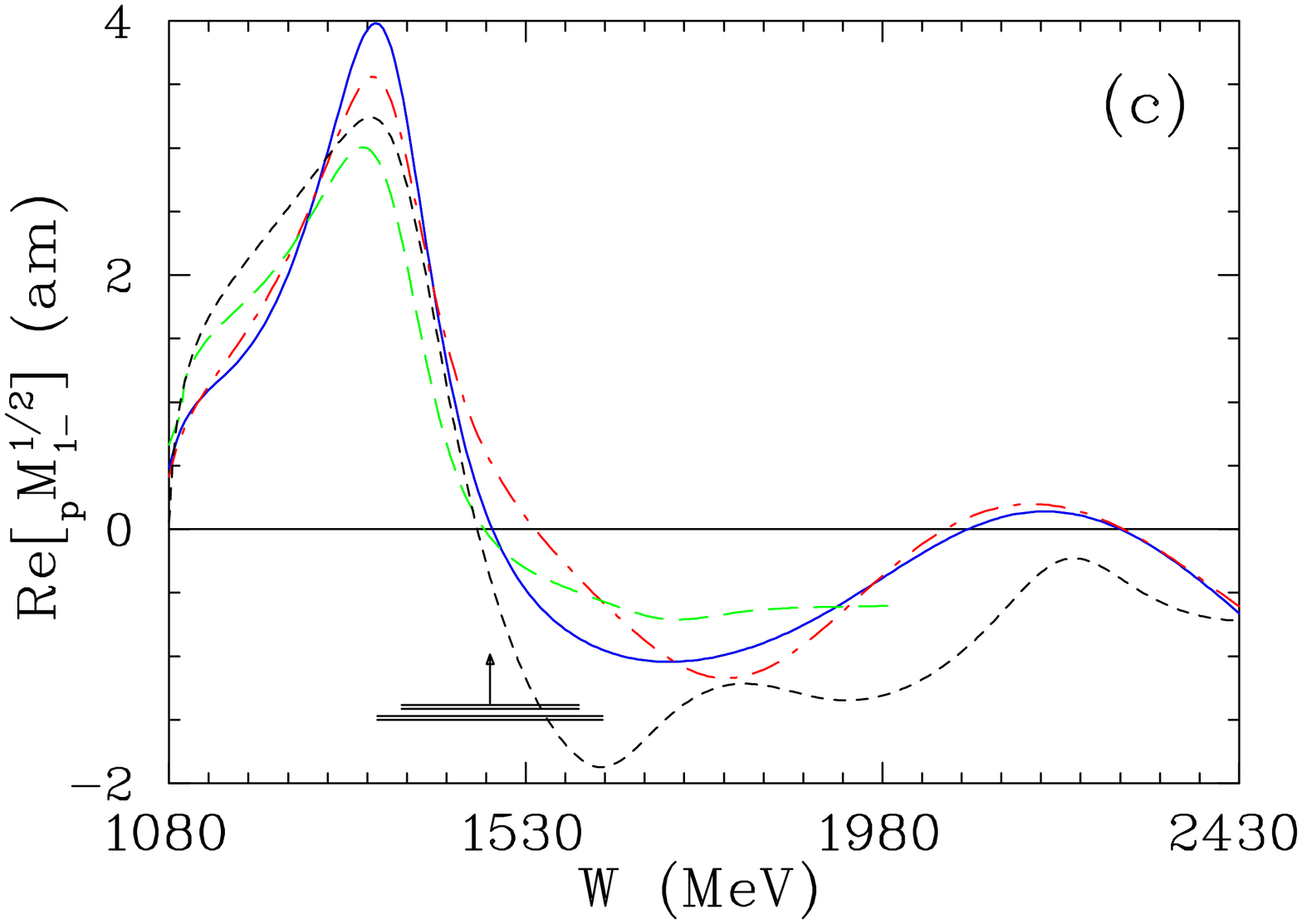}\hfill
\includegraphics[height=0.42\textwidth, angle=90]{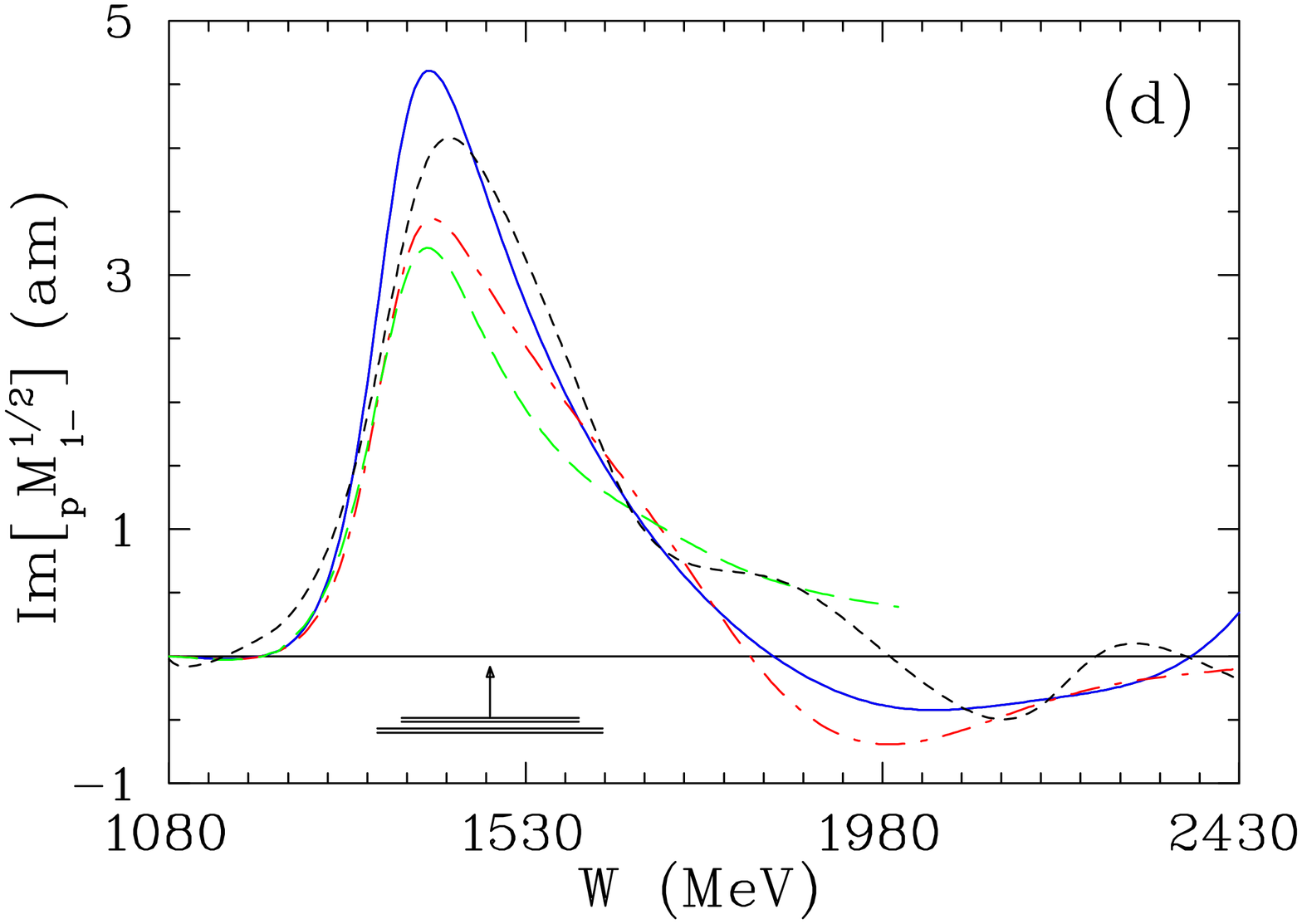}}
\centerline{
\includegraphics[height=0.42\textwidth, angle=90]{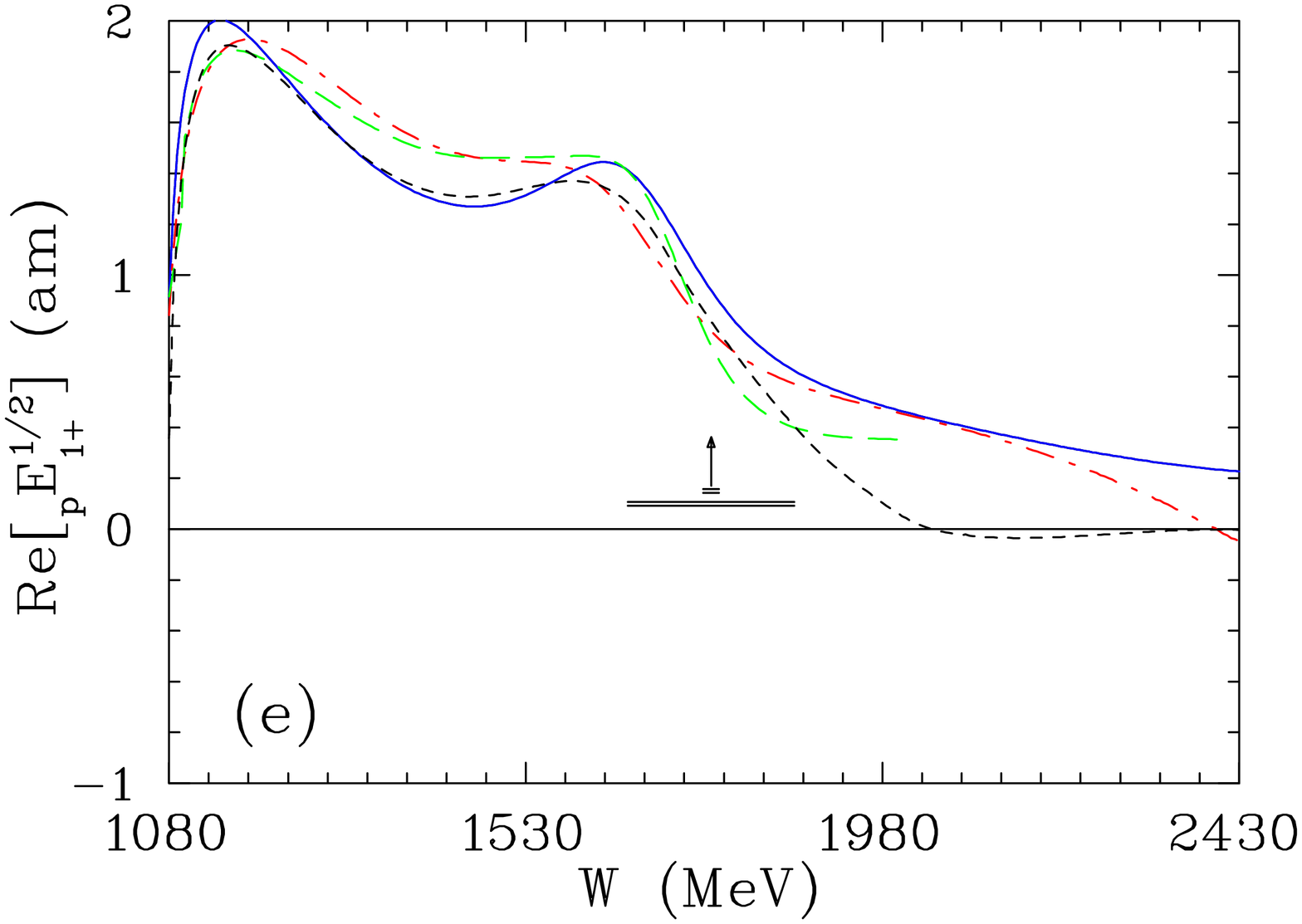}\hfill
\includegraphics[height=0.42\textwidth, angle=90]{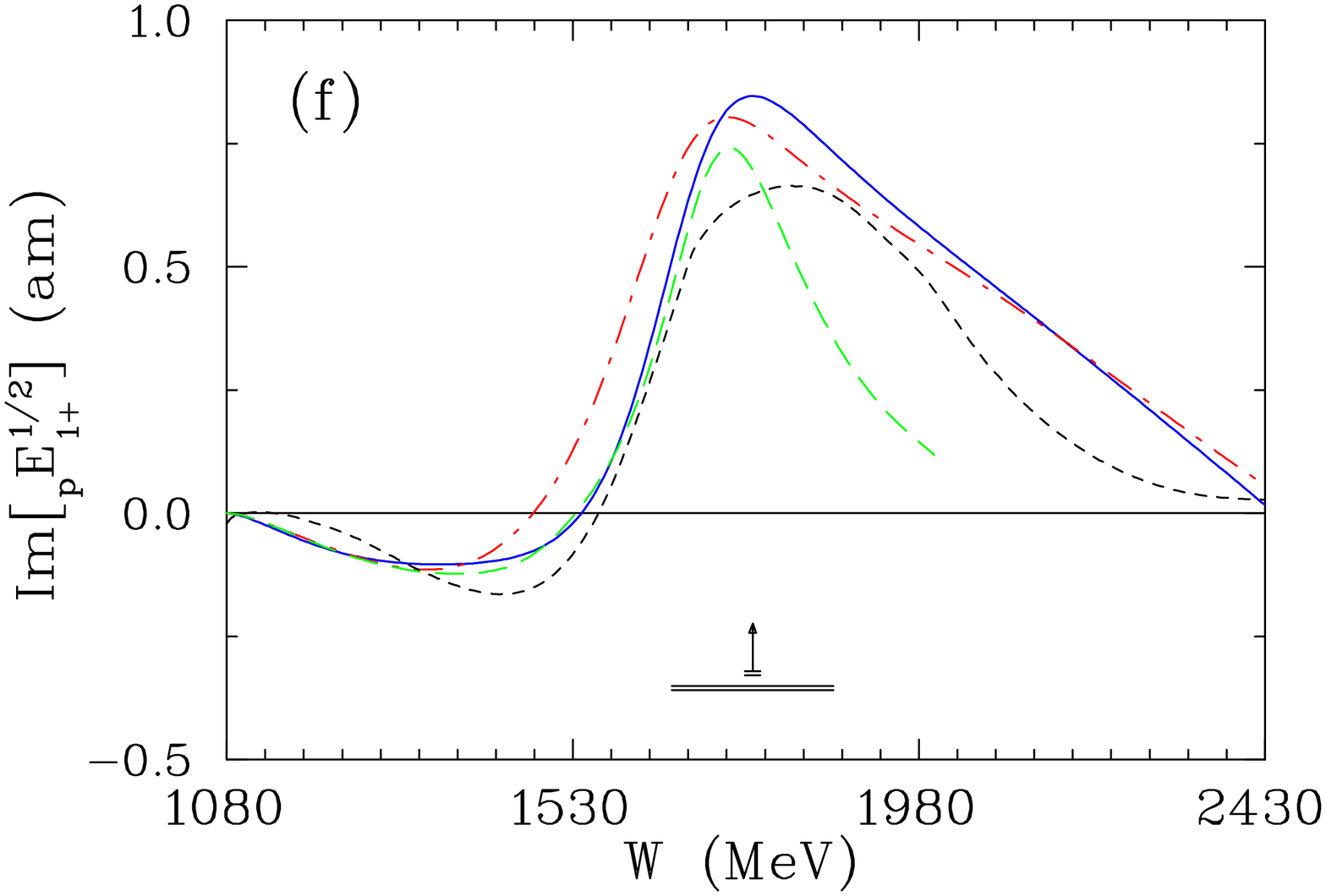}}
\centerline{
\includegraphics[height=0.42\textwidth, angle=90]{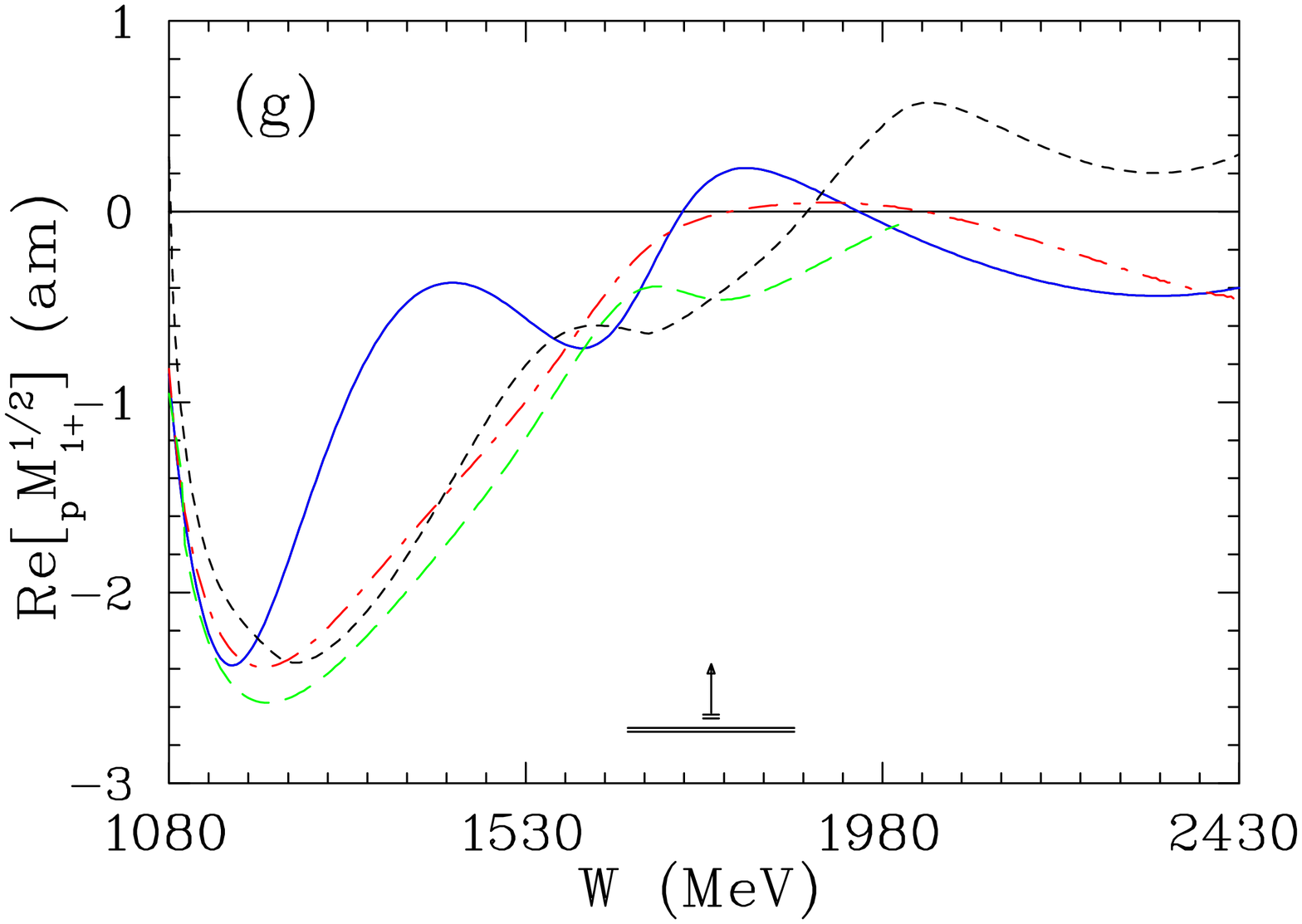}\hfill
\includegraphics[height=0.42\textwidth, angle=90]{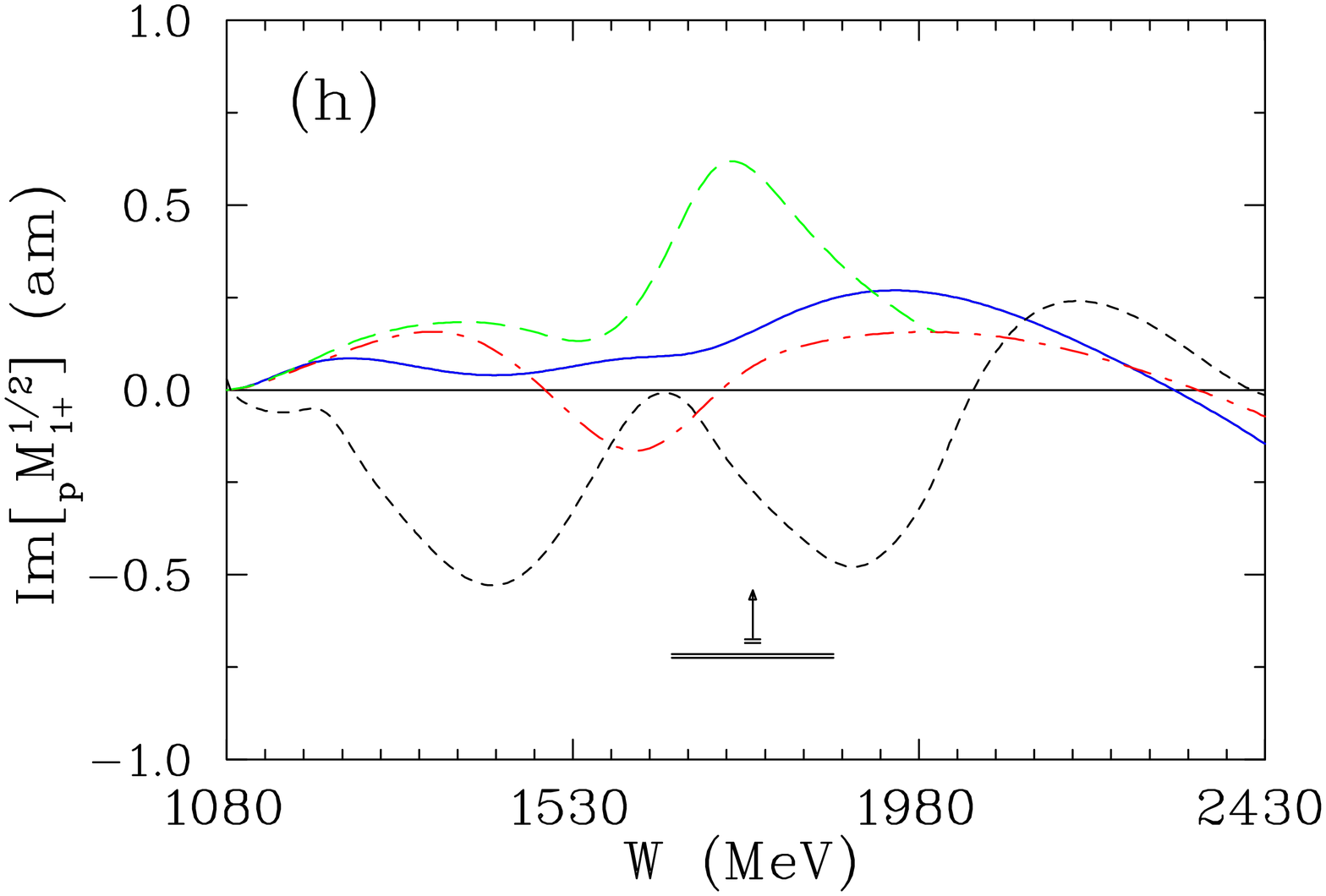}}
\caption{(Color online) Proton multipole I=1/2 amplitudes from 
        threshold to $W$ = 2.43~GeV ($E_{\gamma}$ = 2.7~GeV). 
        Notation of the solutions is the same as in 
        Fig.~\protect\ref{fig:f1}. Vertical arrows indicate 
        resonance energies, $W_R$, and horizontal bars show 
        full ($\Gamma$) and partial ($\Gamma_{\pi N}$) widths 
        associated with the SAID $\pi N$ solution 
        SP06~\protect\cite{Arndt:2006bf}. \label{fig:f4}}
\end{figure*}

\begin{figure*}[th]
\centerline{
\includegraphics[height=0.42\textwidth, angle=90]{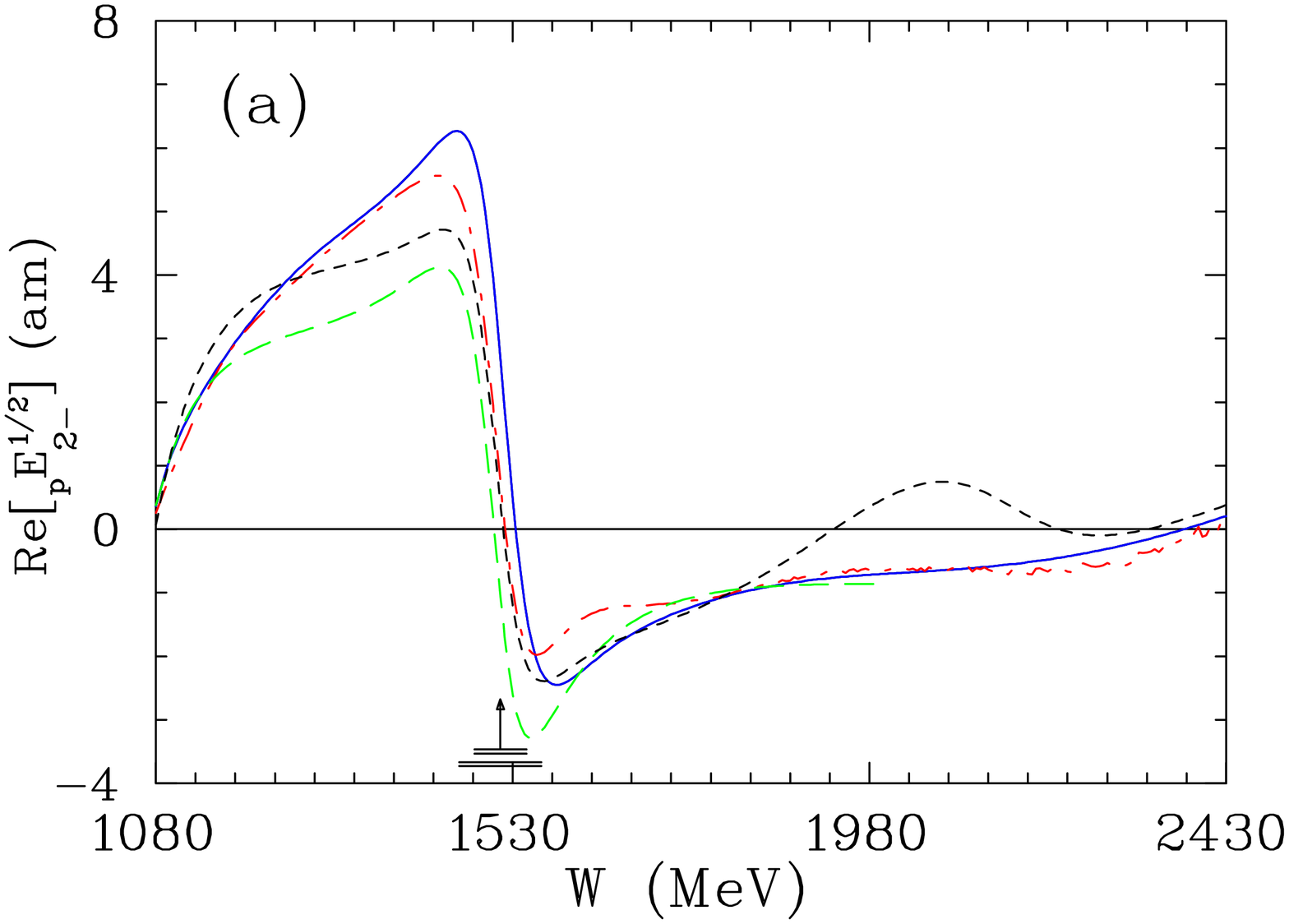}\hfill
\includegraphics[height=0.42\textwidth, angle=90]{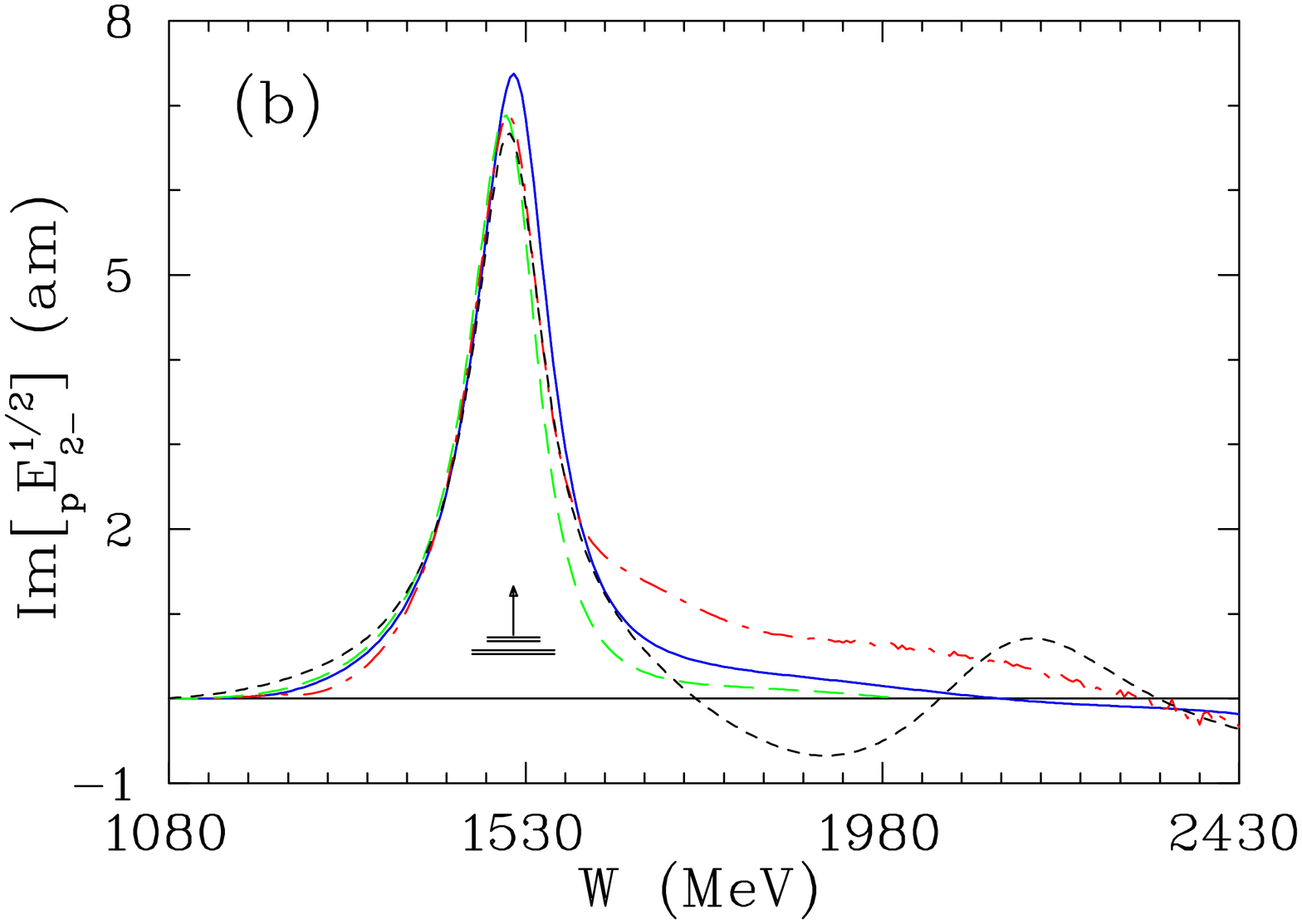}}
\centerline{
\includegraphics[height=0.42\textwidth, angle=90]{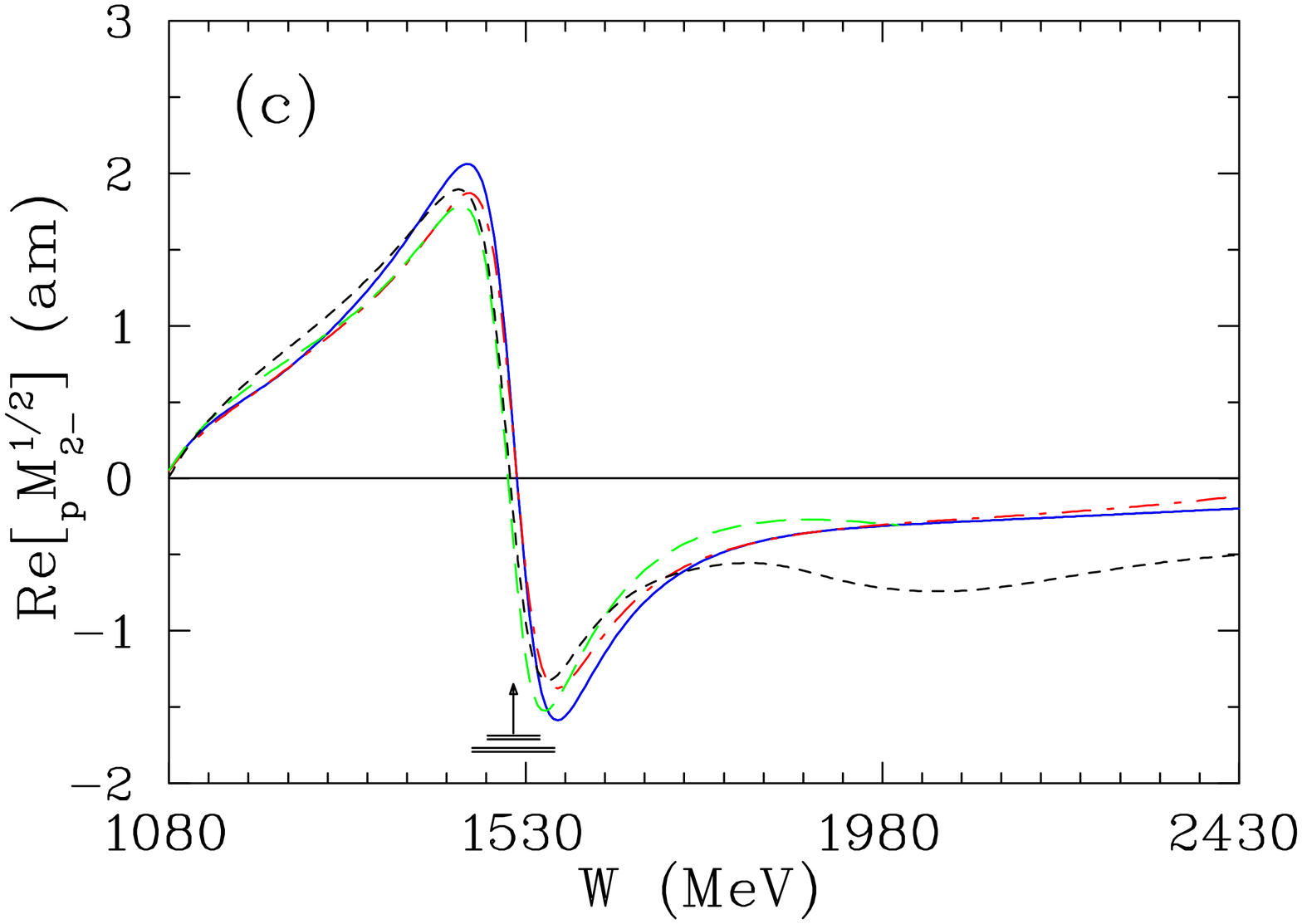}\hfill
\includegraphics[height=0.42\textwidth, angle=90]{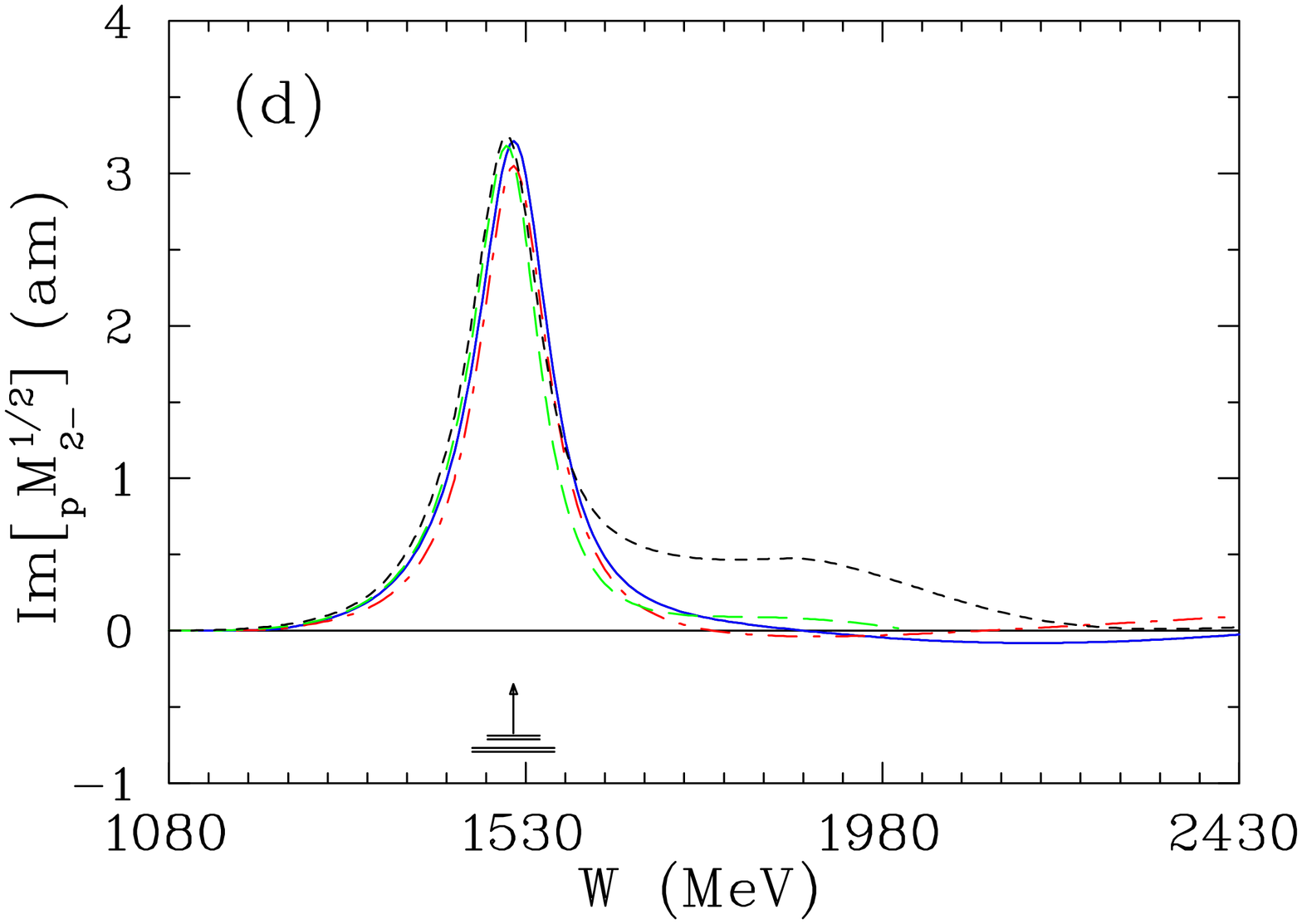}}
\centerline{
\includegraphics[height=0.42\textwidth, angle=90]{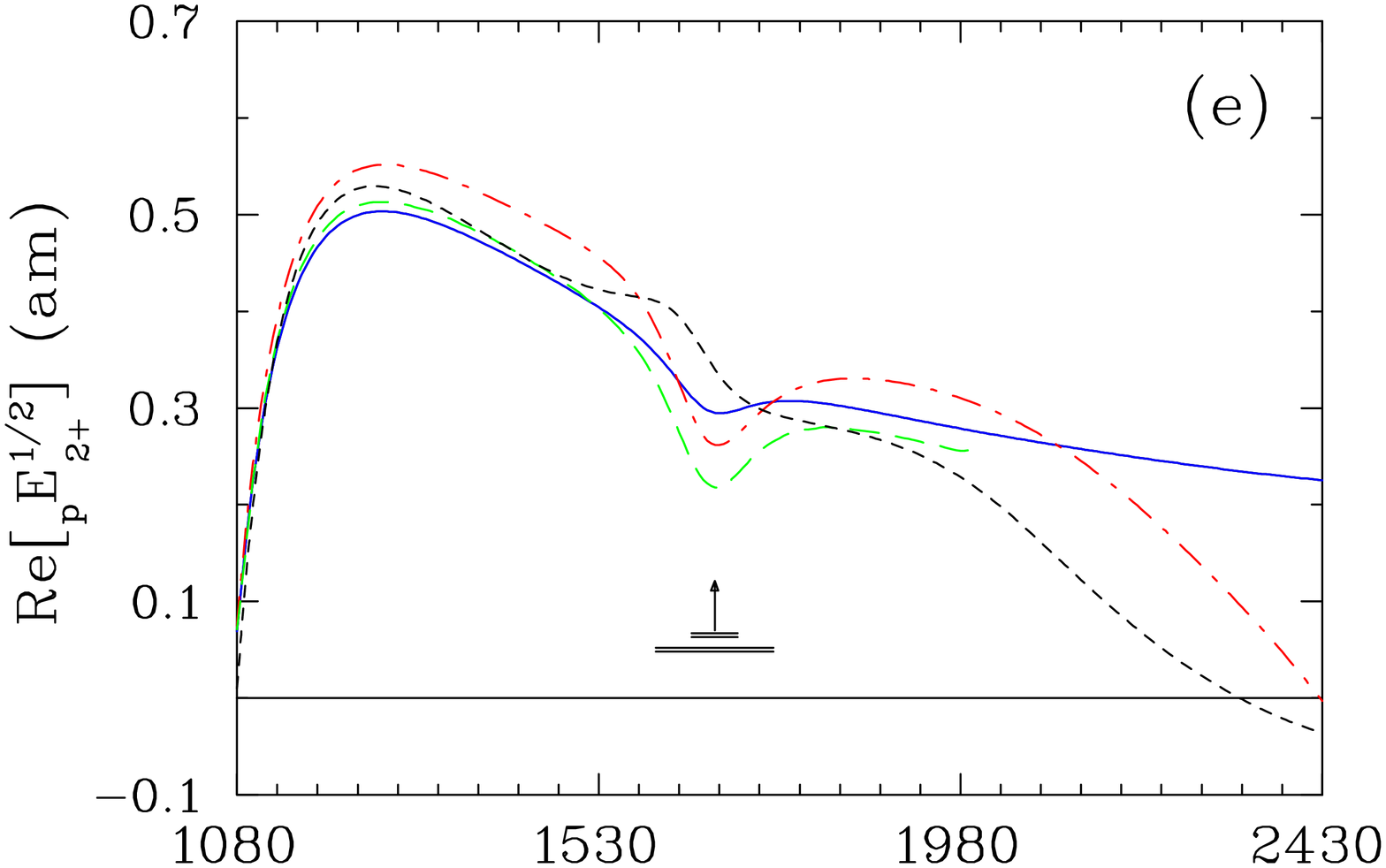}\hfill
\includegraphics[height=0.42\textwidth, angle=90]{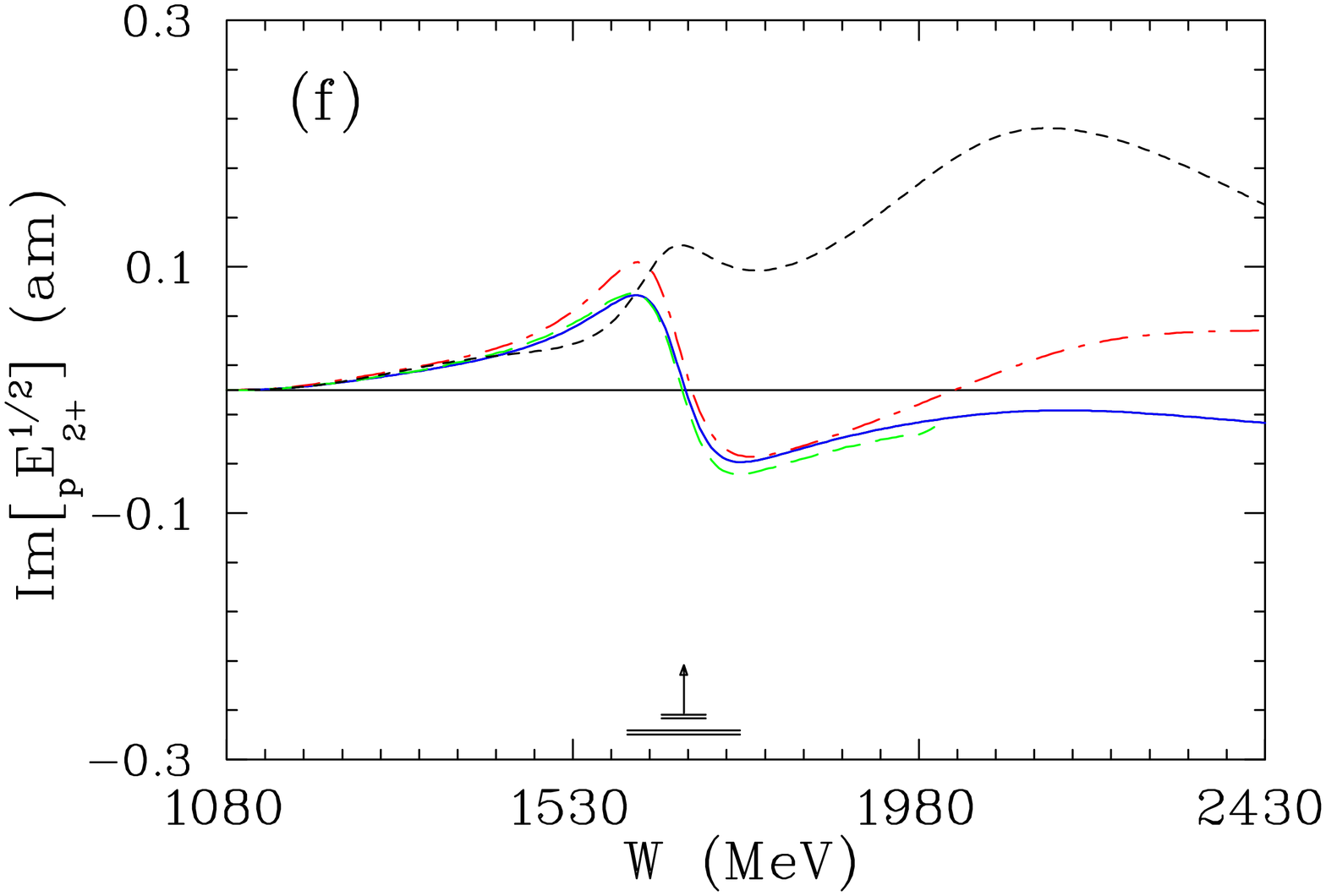}}
\centerline{
\includegraphics[height=0.42\textwidth, angle=90]{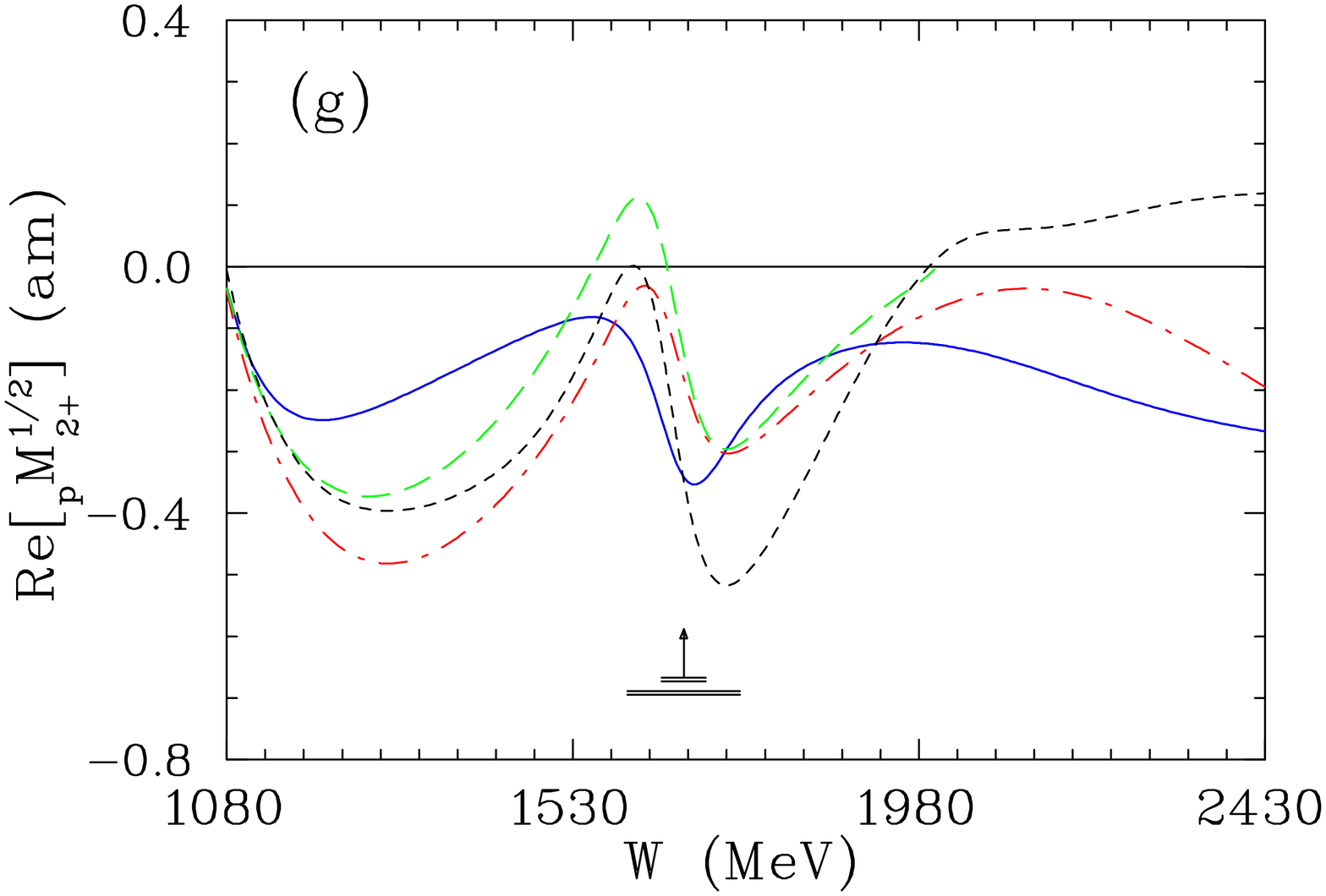}\hfill
\includegraphics[height=0.42\textwidth, angle=90]{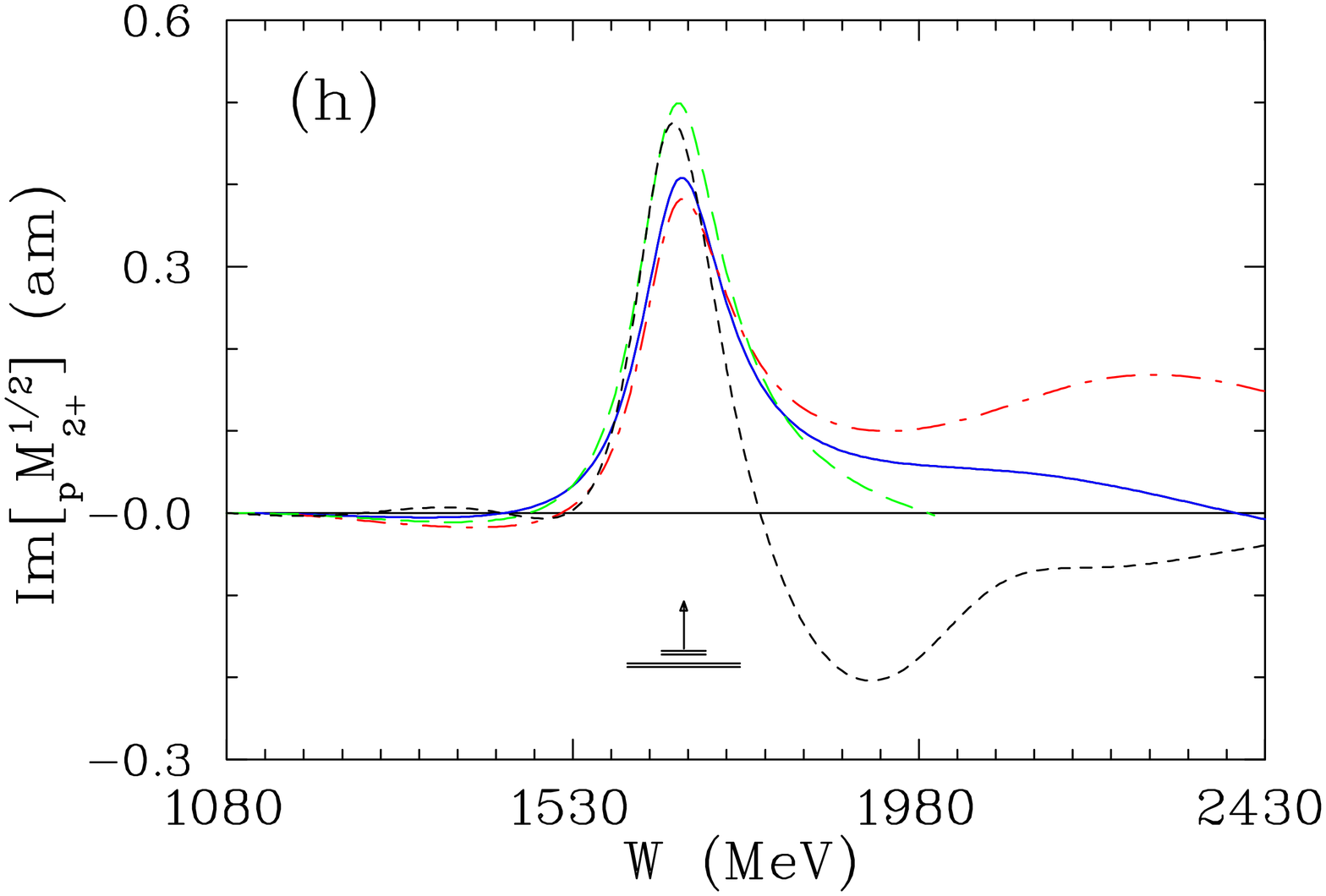}}
\caption{(Color online) Notation of the multipoles is the 
	same as in Fig.~\protect\ref{fig:f4}. \label{fig:f5}}
\end{figure*}

\begin{figure*}[th]
\centerline{
\includegraphics[height=0.42\textwidth, angle=90]{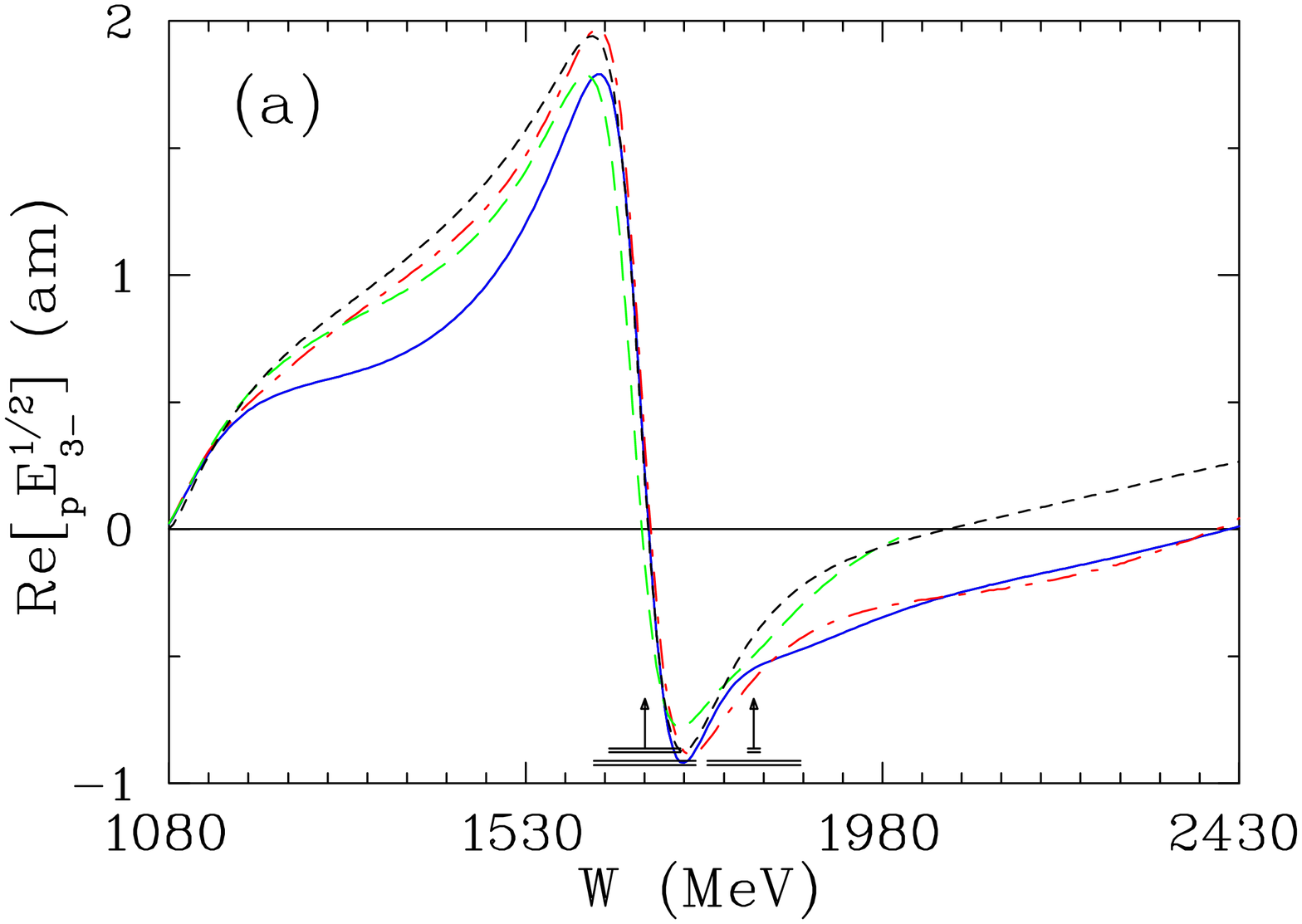}\hfill
\includegraphics[height=0.42\textwidth, angle=90]{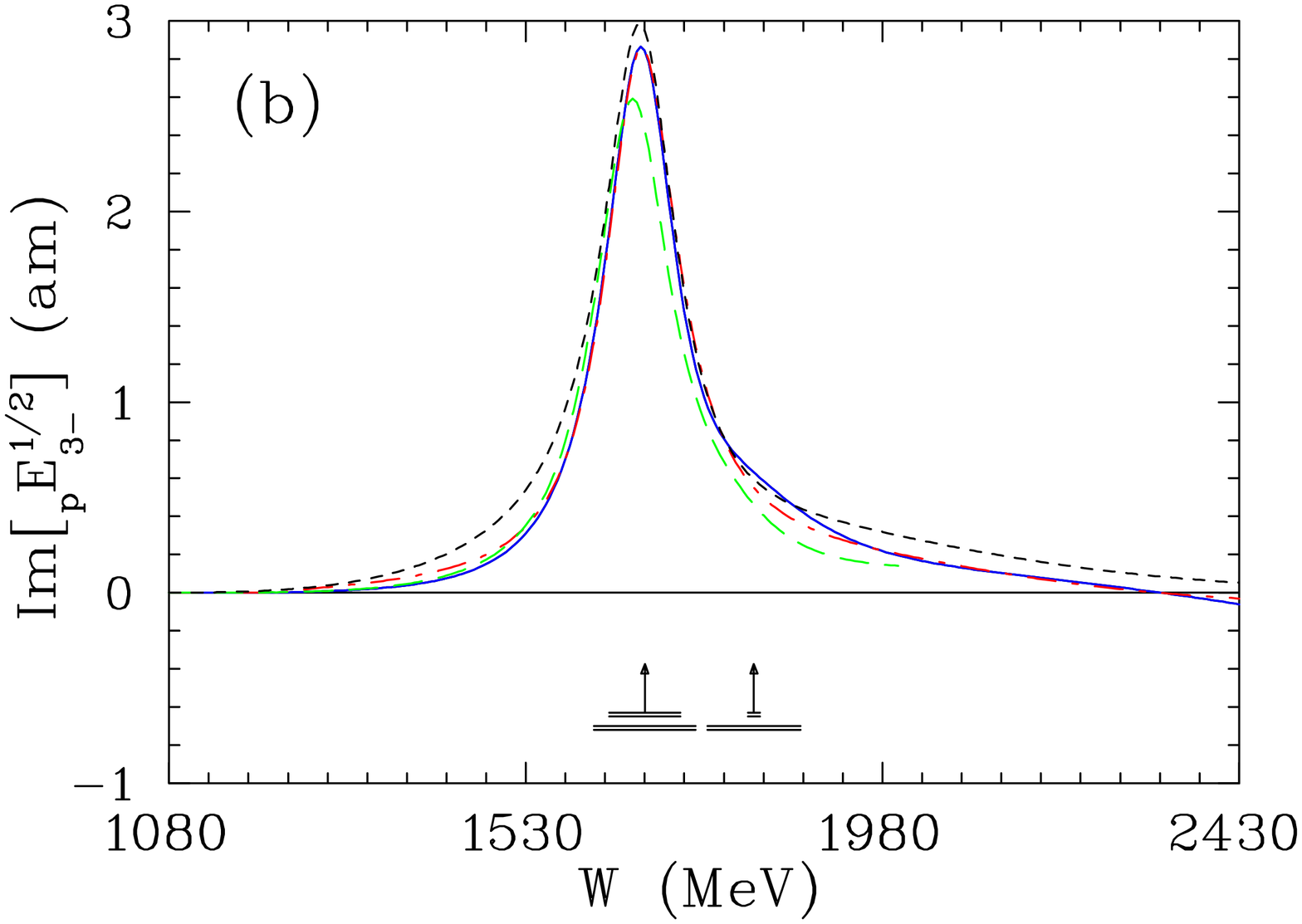}}
\centerline{
\includegraphics[height=0.42\textwidth, angle=90]{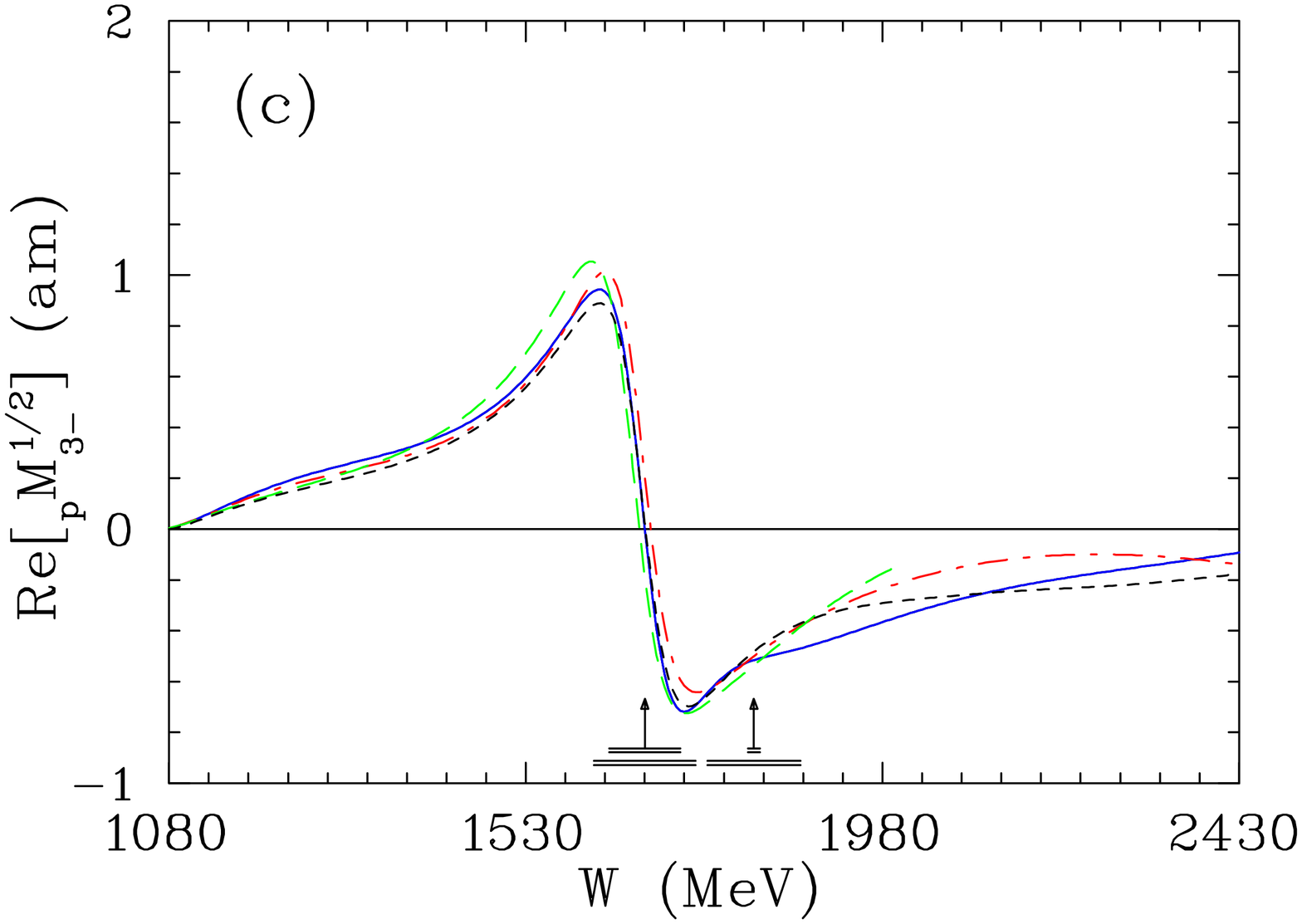}\hfill
\includegraphics[height=0.42\textwidth, angle=90]{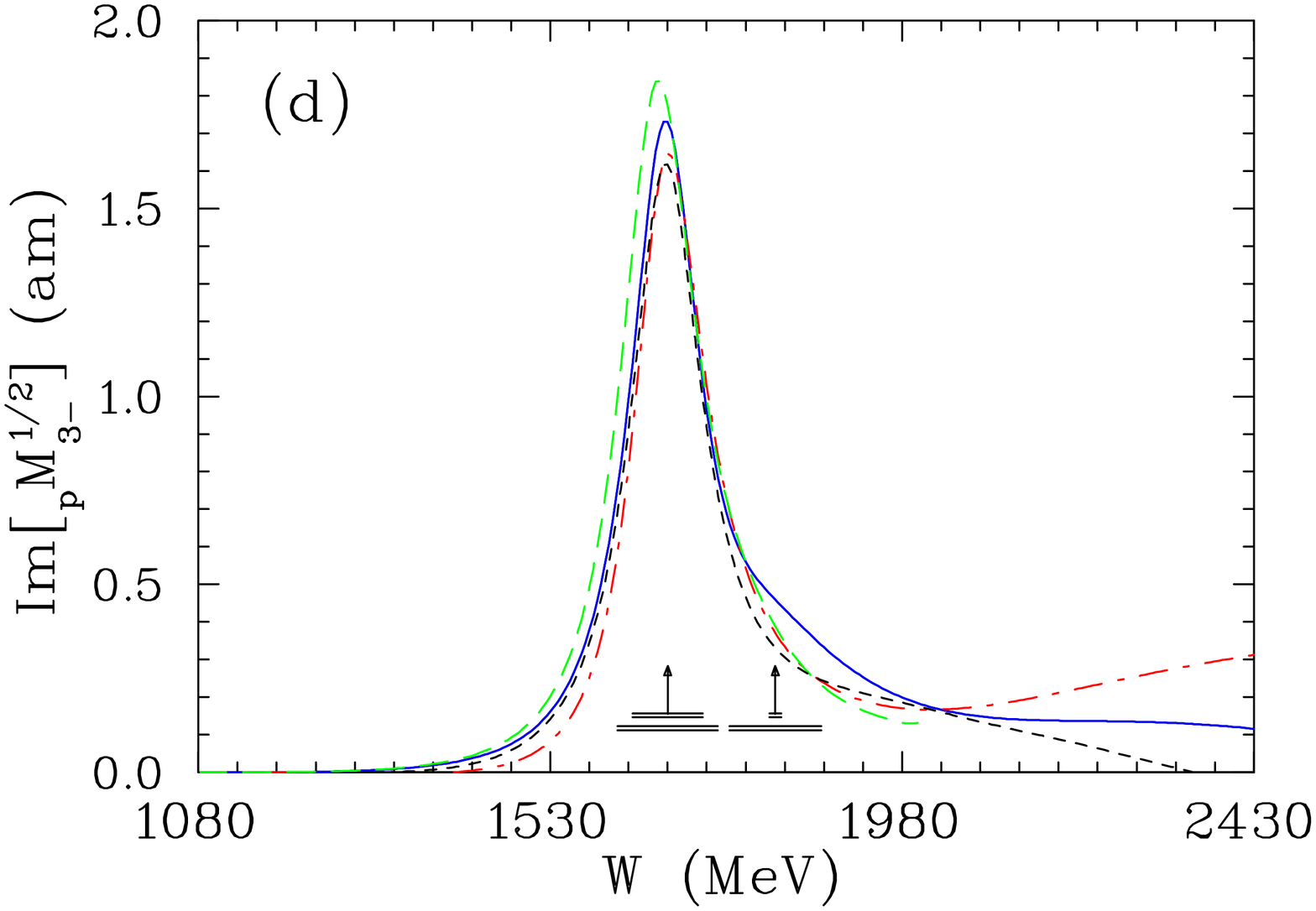}}
\centerline{
\includegraphics[height=0.42\textwidth, angle=90]{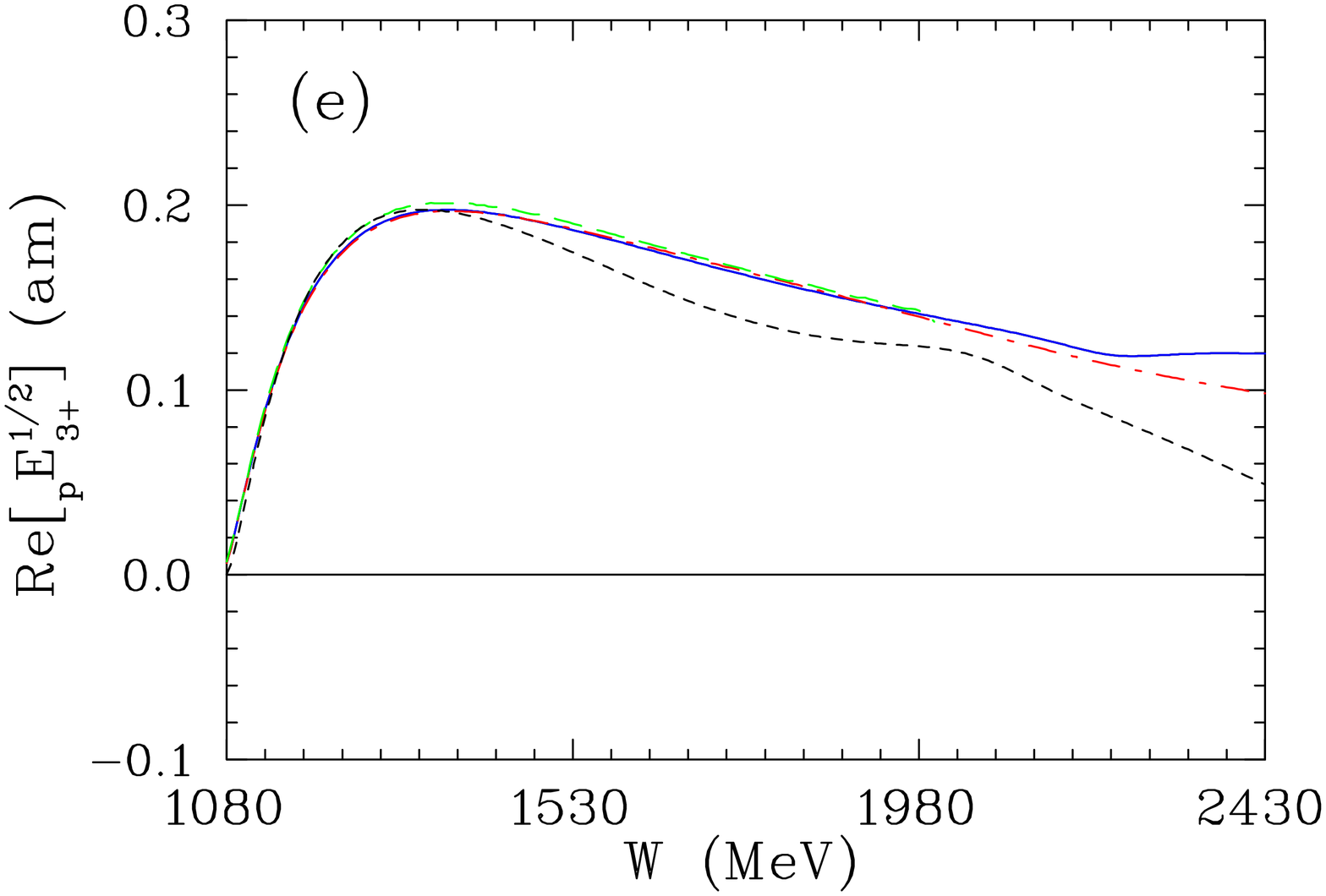}\hfill
\includegraphics[height=0.42\textwidth, angle=90]{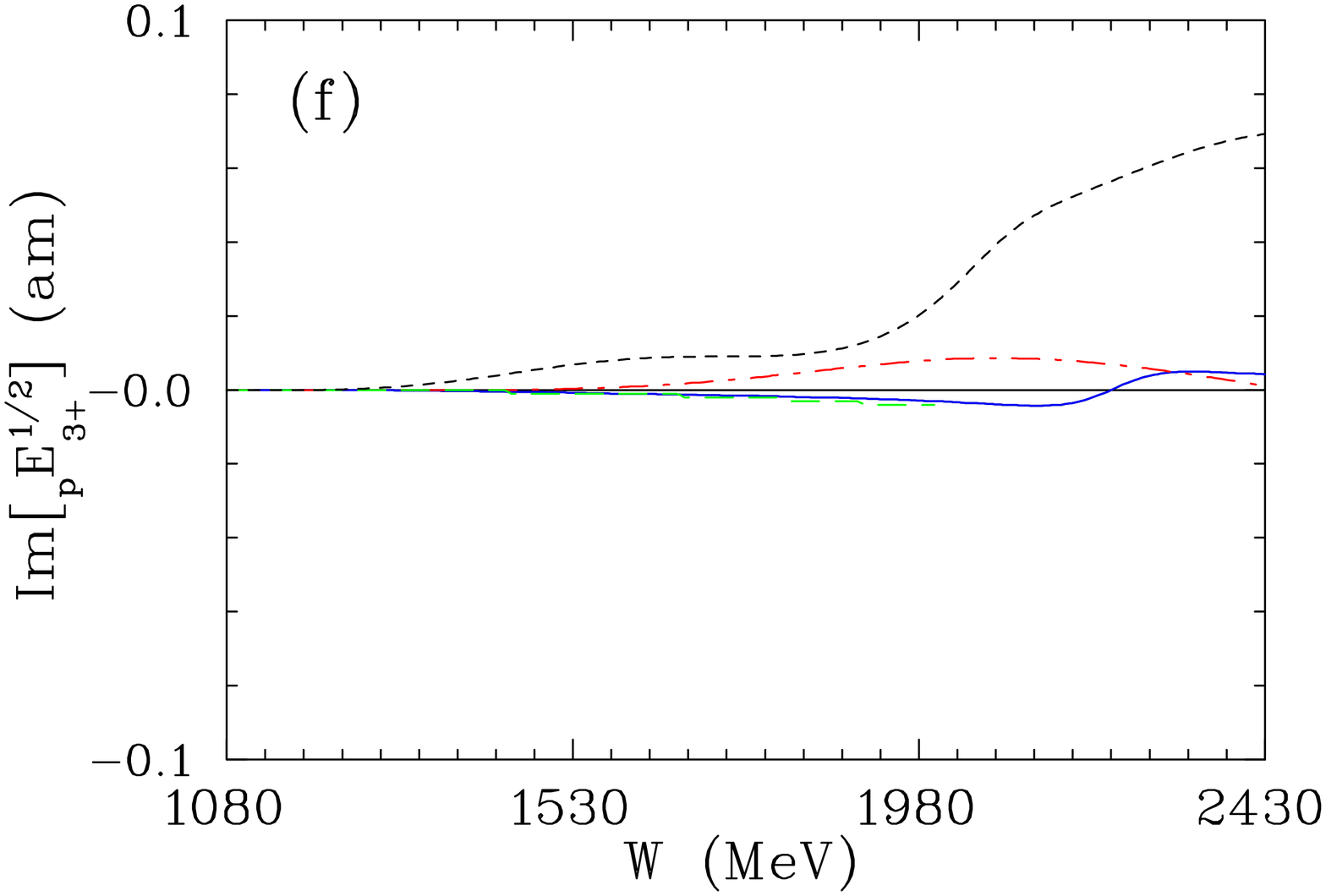}}
\centerline{
\includegraphics[height=0.42\textwidth, angle=90]{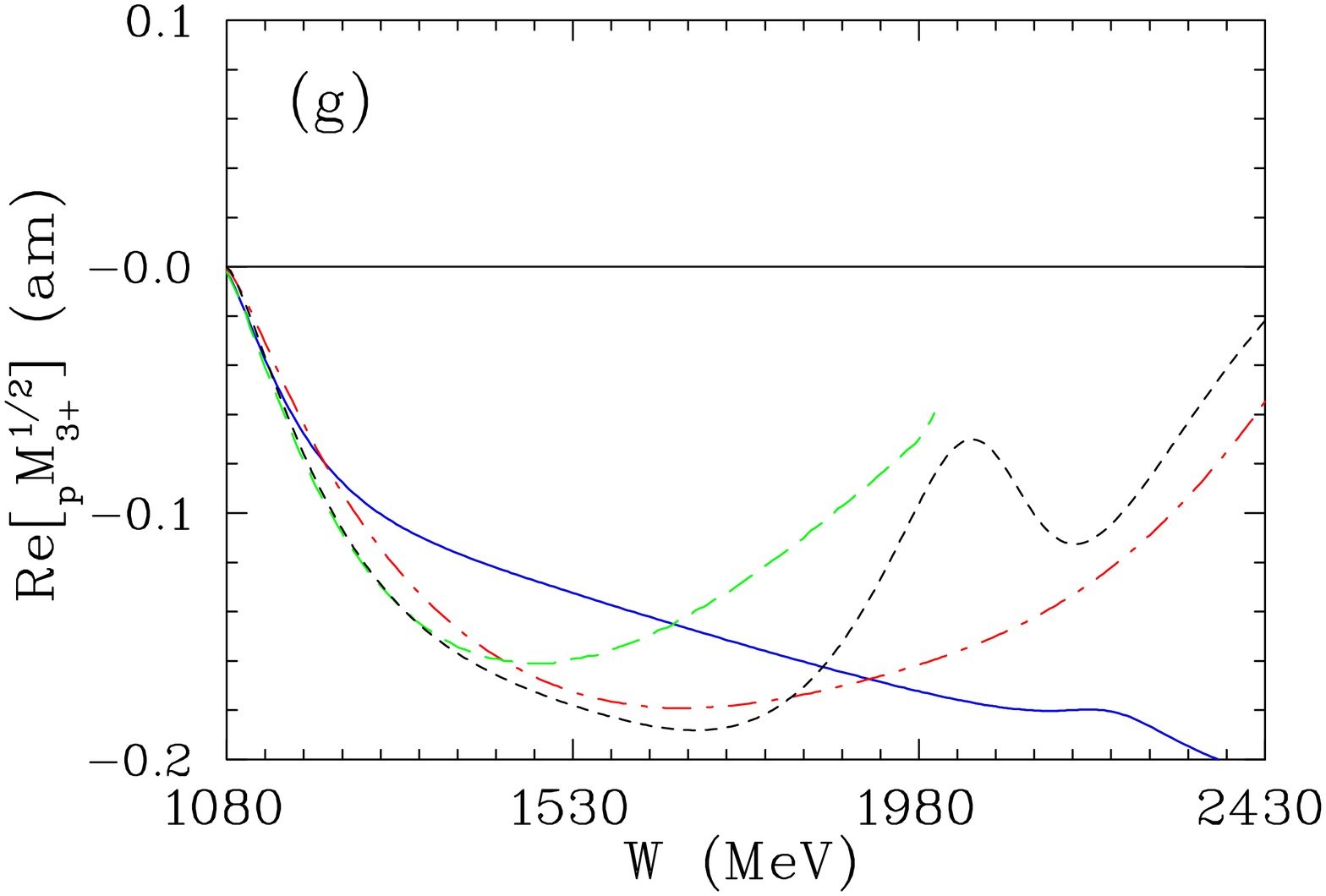}\hfill
\includegraphics[height=0.42\textwidth, angle=90]{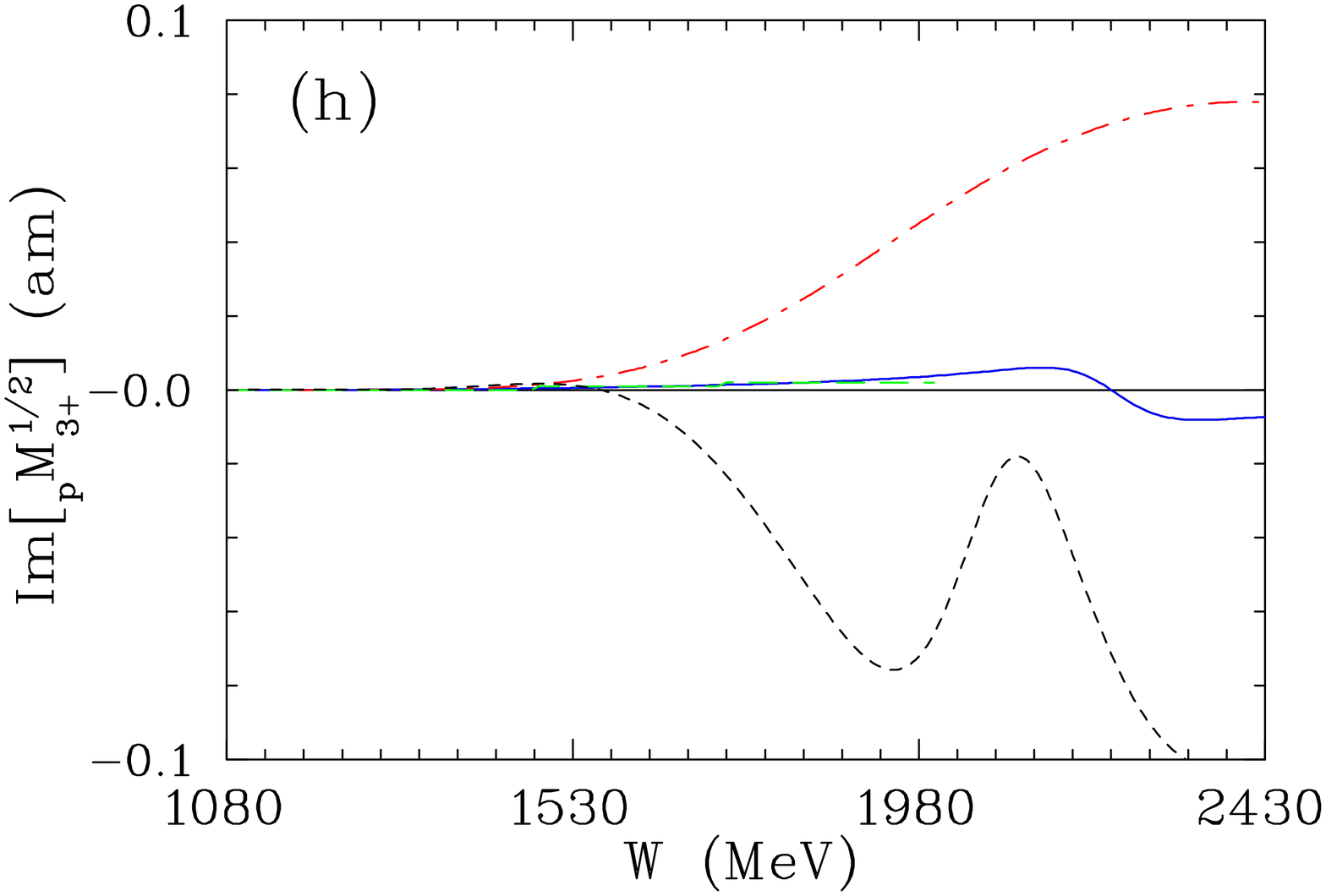}}
\caption{(Color online) Notation of the multipoles is the 
	same as in Fig.~\protect\ref{fig:f4}. \label{fig:f6}}
\end{figure*}

\begin{figure}[th]
\centerline{
\includegraphics[height=0.42\textwidth, angle=90]{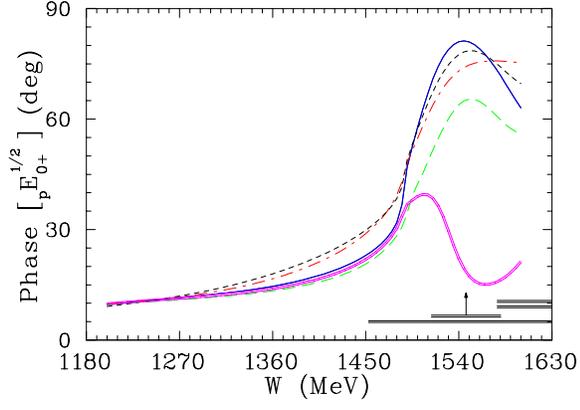}}
\caption{(Color online) Phase for $_pE^{1/2}_{0+}$ multipole. 
	Notation of the solutions is the same as in
        Fig.~\protect\ref{fig:f1}. The thick solid line 
	corresponds to the SAID $\pi N$ solution
        SP06~\protect\cite{Arndt:2006bf} for the
        $S_{11}$ phase shift. \label{fig:f7}}
\end{figure}

In Figures~\ref{fig:f1}$-$\ref{fig:f6}, we compare SN11 and CM12 to 
MAID07~\cite{maid} and Bonn-Gatchina~\cite{BoGa} fits. Only proton-target 
multipoles are presented, as changes in the neutron-target database are 
likely to alter these fits in the near future. For resonances with a 
canonical Breit-Wigner shape, such as the $\Delta (1232) P_{33}$, 
$N(1520)D_{13}$, $N(1680)F_{15}$, and $\Delta(1950)F_{37}$, all solutions 
agree fairly well in the neighborhood of the resonance energy. The fit 
CM12 deviates significantly from SN11 in the $E_{0+}^{1/2}$ multipole. 
The CM12 phase behavior, from threshold up to the peak of the 
$N(1535)S_{11}$ resonance, differs from SN11 and MAID07, but is 
qualitatively similar to the Bonn-Gatchina result, as shown in 
Figure~\ref{fig:f7}. While both SN11 and MAID07 essentially follow the 
$S_{11}$ pion-nucleon phase up to the $\eta N$ threshold cusp, the CM12 
and Bonn-Gatchina fits depart from this phase above the two-pion 
production threshold. 

Some structures occuring in SN11, between threshold and the first 
resonance energies, are missing or diminished in CM12. Examples are the 
real parts of $M_{1-}^{3/2}$, $M_{1+}^{1/2}$, and $E_{3-}^{1/2}$. In each 
of these cases, the CM12 fit more closely resembles the MAID07 and 
Bonn-Gatchina results. This reflects the fact that the replacement of the 
phenomenological terms $A$ and $B$ of Eq.~(\ref{eq:q1}) with a single 
CM~K-matrix element in Eq.~(\ref{eqn:Tag}) is more form-restrictive. 

\begin{figure}[th]
\centerline{
\includegraphics[height=0.42\textwidth, angle=90]{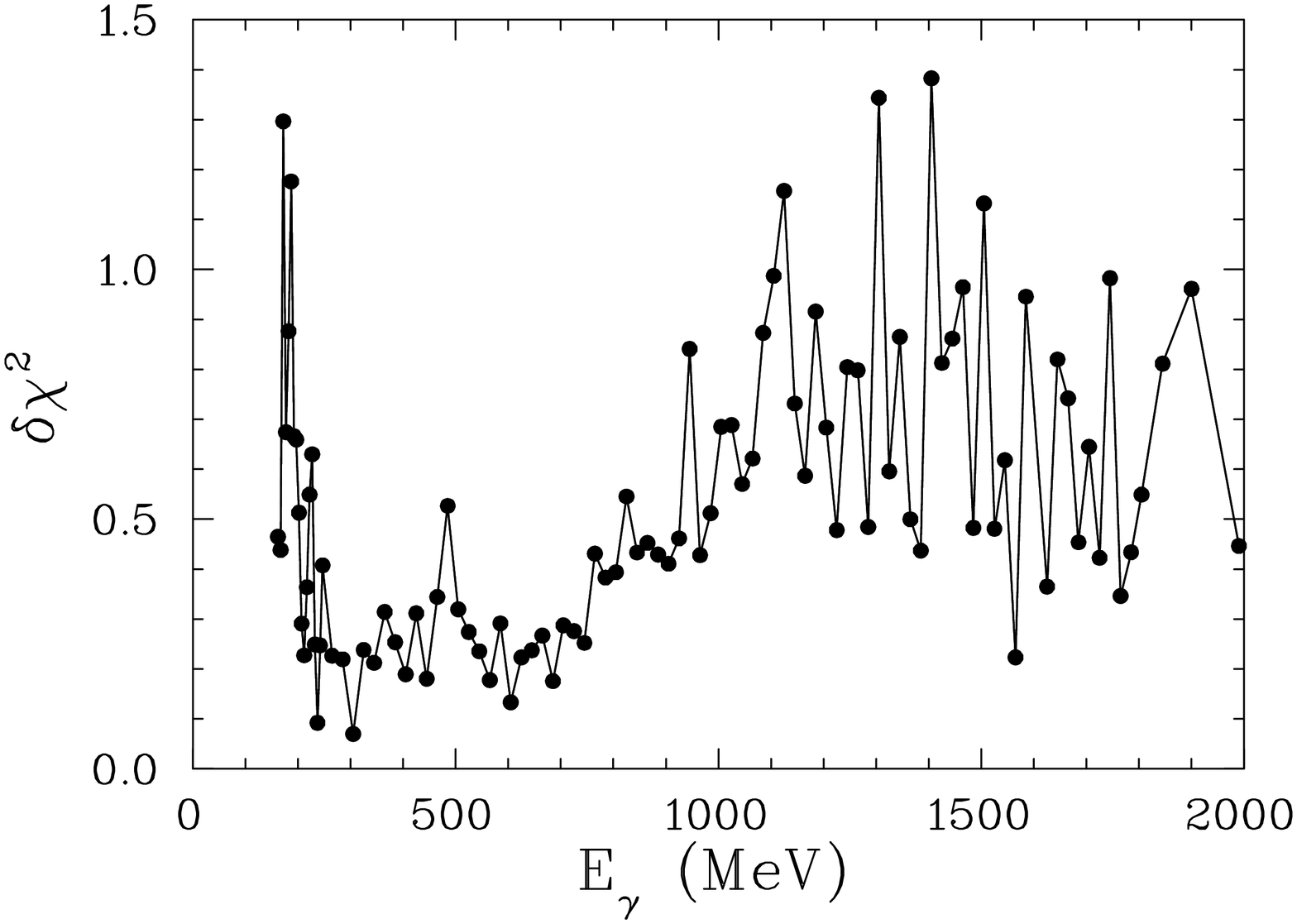}}
\caption{Comparison of the SES and ED CM11 fits via 
	$\delta\chi^2 = {[}\chi^2(CM12)-\chi^2(SES){]}$/N$_{\rm 
	data}$ versus laboratory photon energy $E_{\gamma}$. 
	\label{fig:f8}}
\end{figure}

\begin{figure}[th]
\centerline{
\includegraphics[height=0.42\textwidth, angle=90]{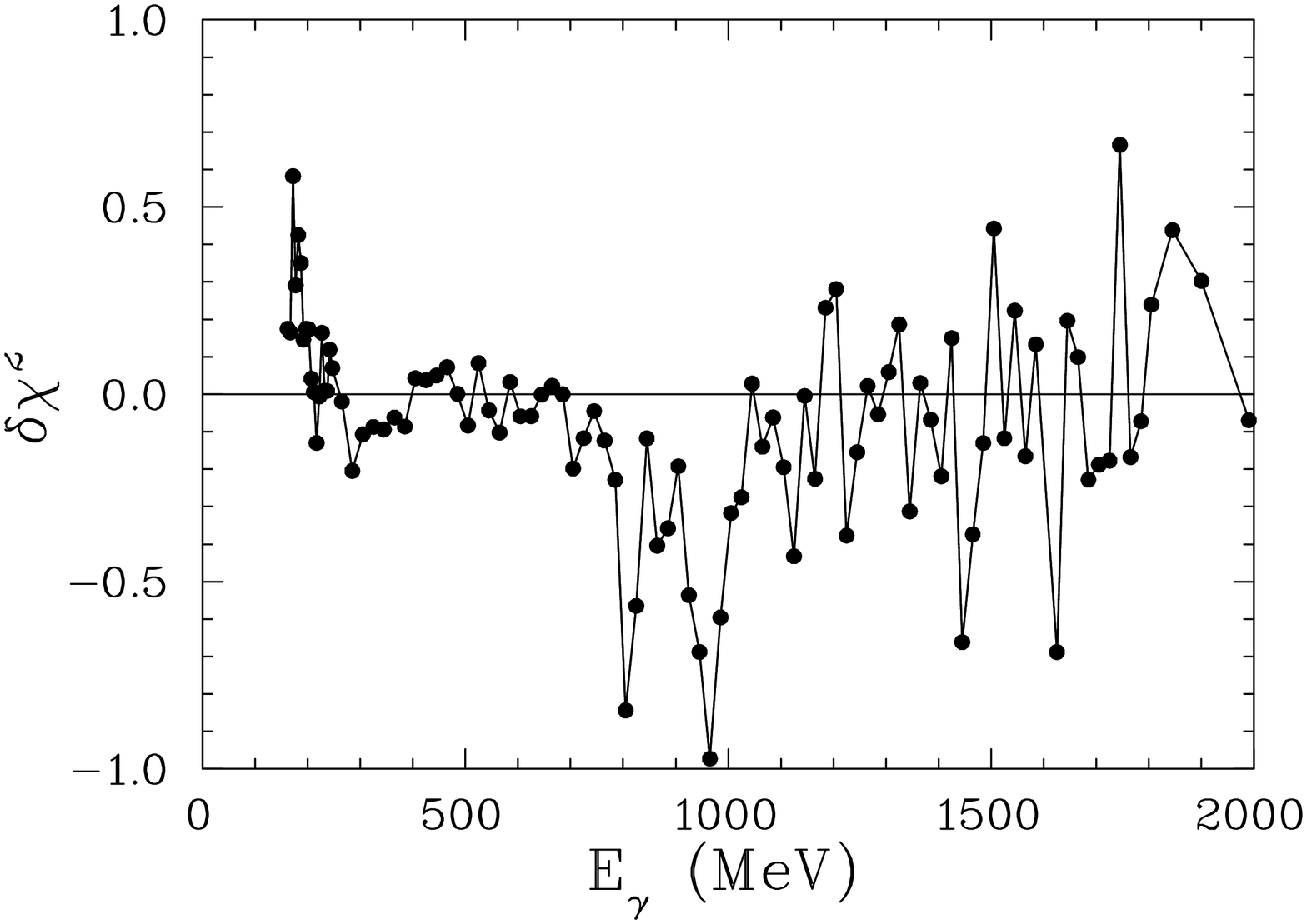}}
\caption{Comparison of the CM12 and SN11 SES fits via $\delta\chi^2 =
        {[}\chi^2(CM12)-\chi^2(SN11){]}$/N$_{\rm data}$ versus
        laboratory photon energy $E_{\gamma}$. \label{fig:f9}}
\end{figure}

In Fig.~\ref{fig:f8}, we compare the $\chi^2$/data of the ED and SES fits
over energy bins used to generate the SES. As in Ref.~\cite{Workman:2011vb}, 
we see a noticeable increase in the $\chi^2$ difference above about 800~MeV. 
In Fig.~\ref{fig:f9}, we compare $\chi^2$ values for SES generated from the 
SN11 and CM12 fits over identical energy bins. As multipole phases are fixed 
for multipoles searched in generating these SES~\cite{Workman:2011hi}, this 
serves as a comparison of the, often quite different, phases found in SN11 
and CM12. While the SN11 SES achieve a better fit in the near-threshold
region, between 700 and 1100~MeV the CM12~SES consistently give the 
better fit. 

%%%%%%%%%%%%%%%%%%%%%%%%%%%%%%%%%%%%%%%%%%%%%%%%%%%%%%%%%%%%%%%%%%%%%%
\begin{table}[th]
\caption{$\chi^2$/data and number of searched parameters ($N_{p}$) 
        compared for fits to pion photoproduction data 
	over varied energy ranges. The fit form used for 
	solution SN11~\protect\cite{Workman:2011vb} is 
	compared to the Chew-Mandelstam form, CM12. See 
	text for details.  \label{tab:tbl1}}
\vspace{2mm}
\begin{tabular}{|c|c|c|c|}
\colrule
Solution & Energy limit& $\chi^2$/N$_{\rm Data}$ & N$_{p}$ \\
         & (MeV)       &               &      \\
\colrule
SN11     & 2700        &  2.08         & 209  \\
CM12     & 2700        &  2.01         & 200 \\
\colrule
SN11a    & 2100        &  1.96         & 206 \\
CM12a    & 2100        &  1.88         & 194 \\
\colrule
SN11b    & 1200        &  1.69         & 175 \\
CM12b    & 1200        &  1.64         & 166 \\
\colrule
\end{tabular}
\end{table}
%%%%%%%%%%%%%%%%%%%%%%%%%%%%%%%%%%%%%%%%%%%%%%%%%%%%%%%%%%%%%%%%

%%%%%%%%%%%%%%%%%%%%%%%%%%%%%%%%%%%%%%%%%%%%%
%%%%%%%%%%%%%%%%%%%%%%%%%%%%%%%%%%%%%%%%%%%%%
%%%%%%%%%%%%%%%%%%%%%%%%%%%%%%%%%%%%%%%%%%%%%
%%%%%%%%%%%%%%%%%%%%%%%%%%%%%%%%%%%%%%%%%%%%%
%%%%%%%%%%%%%%%%%%%%%%%%%%%%%%%%%%%%%%%%%%%%%
%%%%%%%%%%%%%%%%%%%%%%%%%%%%%%%%%%%%%%%%%%%%%
%%%%%%%%%%%%%%%%%%%%%%%%%%%%%%%%%%%%%%%%%%%%%

%%%%%%%%%%%%%%%%%%%%%%%%%%%%%%%%%%%%%%%%%%%%%

%%%%%%%%%%%%%%%%%%%%%%%%%%%%%%%%%%%%%%%%%%%%%

%%%%%%%%%%%%%%%%%%%%%%%%%%%%%%%%%%%%%%%%%%%%%

%%%%%%%%%%%%%%%%%%%%%%%%%%%%%%%%%%%%%%%%%%%%%%%%%%%%%%%%%%%%%%
\begin{table}[th]
\caption{Resonance parameters for N$^\ast$ and $\Delta^\ast$ states
         from the SAID fit to the $\pi N$ data~\protect\cite{Arndt:2006bf}
         (second column) and proton helicity amplitudes $A_{1/2}$ and
         $A_{3/2}$ (in [(GeV)$^{-1/2}\times 10^{-3}$] units) from the 
	 CM12 solution (first row), the SN11~\protect\cite{Workman:2011vb}
         solution (second row), and average values from the
         PDG10~\protect\cite{PDG} (third row). \label{tab:tbl2}}
\vspace{2mm}
\begin{tabular}{|c|c|c|c|}
\colrule
Resonance        & $\pi N$ SAID               &   $A_{1/2}$    & $A_{3/2}$ \\
\colrule
$N(1535)S_{11}$  & $W_{R}$=1547~MeV           &  128$\pm4$    &  \\
                 & $\Gamma$=188~MeV           &   99$\pm$2    &  \\
                 & $\Gamma _{\pi}/\Gamma$=0.36&   90$\pm$30   &  \\
\colrule
$N(1650)S_{11}$  & $W_{R}$=1635~MeV           &   55$\pm$30   &  \\
                 & $\Gamma$=115~MeV           &   65$\pm$25   &  \\
                 & $\Gamma _{\pi}/\Gamma$=1.00&   53$\pm$16   &  \\
\colrule
$N(1440)P_{11}$  & $W_{R}$=1485~MeV           &$-$56$\pm$1    &  \\
                 & $\Gamma$=284~MeV           &$-$58$\pm$1    &  \\
                 & $\Gamma _{\pi}/\Gamma$=0.79&$-$65$\pm$4    &  \\
\colrule
$N(1720)P_{13}$  & $W_{R}$=1764~MeV          &   95$\pm$2     & $-$48$\pm$2  \\
                 & $\Gamma$=210~MeV          &   99$\pm$3     & $-$43$\pm$2  \\
                 & $\Gamma _{\pi}/\Gamma$=0.09&  18$\pm$30    & $-$19$\pm$20 \\
\colrule
$N(1520)D_{13}$  & $W_{R}$=1515~MeV          &$-$19$\pm$2     & 141$\pm$2    \\
                 & $\Gamma$=104~MeV          &$-$16$\pm$2     & 156$\pm$2    \\
                 & $\Gamma _{\pi}/\Gamma$=0.63&$-$24$\pm$9    & 166$\pm$5    \\
\colrule
$N(1675)D_{15}$  & $W_{R}$=1674~MeV          &   13$\pm$1     &  16$\pm$1    \\
                 & $\Gamma$=147~MeV          &   13$\pm$2     &  19$\pm$2    \\
                 & $\Gamma _{\pi}/\Gamma$=0.39&  19$\pm$8     &  15$\pm$9    \\
\colrule
$N(1680)F_{15}$  & $W_{R}$=1680~MeV          &$-$7$\pm$2      & 140$\pm$2    \\
                 & $\Gamma$=128~MeV          &$-$13$\pm$3     & 141$\pm$3    \\
                 & $\Gamma _{\pi}/\Gamma$=0.70&$-$15$\pm$6    & 133$\pm$12   \\
\colrule
\colrule
$\Delta(1620)S_{31}$& $W_{R}$=1615~MeV       &   29$\pm$3     & \\
                 & $\Gamma$=147~MeV          &   64$\pm$2     & \\
                 & $\Gamma _{\pi}/\Gamma$=0.32&  27$\pm$11    & \\
\colrule
$\Delta(1232)P_{33}$& $W_{R}$=1233~MeV       &$-$139$\pm$2    &$-$262$\pm$3  \\
                 & $\Gamma$=119~MeV          &$-$138$\pm$3    &$-$259$\pm$5  \\
                 & $\Gamma _{\pi}/\Gamma$=1.00&$-$135$\pm$6   &$-$250$\pm$8  \\
\colrule
$\Delta(1700)D_{33}$& $W_{R}$=1695~MeV       &  105$\pm$5     & 92$\pm$4     \\
                 & $\Gamma$=376~MeV          &  109$\pm$4     & 84$\pm$2     \\
                 & $\Gamma _{\pi}/\Gamma$=0.16& 104$\pm$15    & 85$\pm$22    \\
\colrule
$\Delta(1905)F_{35}$& $W_{R}$=1858~MeV       &    19$\pm$2    &$-$38$\pm$4   \\
                 & $\Gamma$=321~MeV          &    9$\pm$3     &$-$46$\pm$3   \\
                 & $\Gamma _{\pi}/\Gamma$=0.12&   26$\pm$11   &$-$45$\pm$20  \\
\colrule
$\Delta(1950)F_{37}$& $W_{R}$=1921~MeV       & $-$83$\pm$4    &$-$96$\pm$4  \\
                 & $\Gamma$=271~MeV          & $-$71$\pm$2    &$-$92$\pm$2   \\
                 & $\Gamma _{\pi}/\Gamma$=0.47& $-$76$\pm$12  &$-$97$\pm$10  \\
\colrule
\end{tabular}
\end{table}
%%%%%%%%%%%%%%%%%%%%%%%%%%%%%%%%%%%%%%%%%%%%%%%%%%%%%%%%%%%%%

%%%%%%%%%%%%%%%%%%%%%%%%%%%%%%%%%%%%%%%%%%%%%%%%%%%%%%%%%%%%%%
\section{Summary and Conclusion}
\label{sec:sum}

We have fitted the single-pion photoproduction database utilizing a parametrization
consistent with the Chew-Mandelstam form used in our previous fits
to $\pi N$ scattering and $\eta N$ production data. This new fit has
a number of interesting features. It is more economical, using fewer
parameters to obtain a slightly better overall fit. Some low-energy
structures, not seen by other groups, have disappeared in the present
fit. The phase behavior of the $E_{0+}^{1/2}$ has changed significantly 
and now is qualitatively similar to the Bonn-Gatchina result. 

Comparison of the ED and SES fits shows, as was found in Ref.~\cite{Workman:2011vb},
a rise in $\chi^2$ difference, evaluated over narrow energy bins, above
about 800 MeV in the photon energy. This could be related to the energy 
limit of MAMI-B, which has contributed a significant fraction of the 
precise data below 800 MeV (thus, a data issue) or due to the treatment
of channels above single-pion production (a model issue). The comparison
of SES fits, derived from the ED SN11 and CM12 solutions, suggests that the
CM12 multipole phases, held fixed in SES, are prefered in the intermediate-energy
region. 

In Table~\ref{tab:tbl2}, we compare photo-decay couplings extracted
from CM12 and SN11, using the method of Ref.~\cite{Workman:2011vb},
to the average of PDG values. As expected, the $N(1535)S_{11}$ shows
a large increase to 128$\pm$4 (in GeV)$^{-1/2}\times 10^{-3}$ units),
compared to 105$\pm$10 from the Bonn-Gatchina group, and 
118 found in Eta-MAID~\cite{etam}. 
As in the SN11 analysis, the $N(1650)S_{11}$ is very difficult to 
fit using this procedure. The fit prefers a larger Breit-Wigner 
mass and width, and a lower value for $\Gamma_{\pi}/ \Gamma$. Allowing
for these variations, as reflected in its uncertainty, the value 
quoted here should be taken only as a rough estimate.

Other couplings are generally 
either close to values found in SN11, or are within the PDG ranges. 
The $N(1520)D_{13}$ $A_{3/2}$ (141$\pm$2) has dropped below the PDG
range (166$\pm$5), but is above the Bonn-Gatchina value of 131$\pm$10.
The $N(1720)P_{13}$ couplings remain very uncertain, as is clear from
the multipole plot comparison. Here the present PDG range is certainly too
narrow - the Bonn-Gatchina result for $A_{3/2}$ is 150$\pm$30, compared
to the PDG range $-19\pm 20$. For $A_{1/2}$, the PDG range is $18\pm 30$, with
the CM12, SN11, and Bonn-Gatchina results all near 100.

%%%%%%%%%%%%%%%%%%%%%%%%%%%%%%%%%%%%%%%%%%%%%%%%%%%%%%%%%%%%%%
\begin{acknowledgments}
This work was supported in part by the U.S.\ Department of Energy
Grant DE-FG02-99ER41110. 
\end{acknowledgments}

%%%%%%%%%%%%%%%%%%%%%%%%%%%%%%%%%%%%%%%%%%%%%%%%%%%%%%%%%%%%%%
%\bibliography{master}

\begin{thebibliography}{99}
\bibitem{limits} For multipoles with $\ell \ge 4$, the new form offers
        no advantage and unitarized Born terms were generally used. The 
        $\Delta (1232)$, associated with a $\pi N$ amplitude 
        which remains elastic to much higher energies,
        is unaltered in the new fit.
\end{thebibliography}

%%%%%%%%%%%%%%%%%%%%%%%%%%%%%%%%%%%%%%%%%%%%%%%%%%%%%%%%%%%%%%
\end{document}